\documentclass[a4paper,11pt]{article}
\usepackage{amsmath}
\usepackage{jheppub} 
\usepackage{lineno}
\usepackage{float}
\usepackage{tikz}
\usepackage{subcaption}
\usepackage{comment}
\usepackage{enumitem}  
\usetikzlibrary{positioning}
\usepackage[compat=1.1.0]{tikz-feynman}
\tikzfeynmanset{warn luatex=false}
\graphicspath{{../figures_paper_ALP/}{./}}

\title{Searching for Dark Photons and ALPs using Gamma-rays from Nuclear Reactors}

\author[a,b,c]{Lucia Angel}
\author[c]{Manfred Lindner}
\author[a,c,d,e]{Farinaldo S. Queiroz}
\author[c]{Bilge S. Solak}

\affiliation[a]{International Institute of Physics, Universidade Federal do Rio Grande do Norte, Campus Universitario, Lagoa Nova, Natal-RN 59078-970, Brazil}
\affiliation[b]{Department of Physics, University of Oslo, Box 1048, N-0316 Oslo, Norway}
\affiliation[c]{Max Planck Institut fur Kernphysik, Heidelberg, Germany}
\affiliation[d]{Millennium Institute for Subatomic Physics at the High-Energy Frontier (SAPHIR) of ANID, Fernández Concha 700, Santiago, Chile}
\affiliation[e]{Departamento de Fisica, Facultad de Ciencias, Universidad de La Serena, Avenida Cisternas 1200,La Serena, Chile}

\emailAdd{lucia.correa.717@ufrn.edu.br}
\emailAdd{lindner@mpi-hd.mpg.de}
\emailAdd{farinaldo.queiroz@ufrn.edu.br}
\emailAdd{solak@mpi-hd.mpg.de}

\begin{document}

\abstract{\noindent Nuclear reactors are intense sources of MeV gamma rays, produced in fission, radiative neutron capture and the de-excitation of fission products.  Building on earlier phenomenological studies, we present a detector-independent description of the
production of light, weakly coupled particles by reactor gamma rays and of their detection through scattering, absorption and decay in a nearby
detector.  We consider kinetically mixed dark photons, produced and detected through Compton-like processes, and axion-like particles (ALPs) coupled to photons, electrons and nucleons, produced through Primakoff conversion, Compton-like scattering and magnetic nuclear transitions such as the $2.23~\mathrm{MeV}$ line of $p(n,\gamma)d$.  As a concrete application, we derive the projected sensitivities of the CONUS+ germanium experiment at the Leibstadt reactor and of a conceivable $100$-kg CONUS100 upgrade.  For sub-MeV dark photons, CONUS+ and CONUS100 reach kinetic mixings of $\epsilon\simeq1.65\times10^{-5}$ and $1.2\times10^{-5}$, respectively, for $0.5~\mathrm{keV}\lesssim m_{A'}\lesssim0.5~\mathrm{MeV}$. 
For ALPs, we show that CONUS100 can probe an unexplored region of parameter space at MeV masses, in particular through $a\to e^+e^-$ decays of ALPs emitted in the deuteron line, assuming a energy window extended to $10~\mathrm{MeV}$.}

\maketitle
\flushbottom

\section{Introduction}
\label{Intro}

A wide range of astrophysical and cosmological observations indicates that
most of the matter content of the Universe is non-luminous and cannot be
accounted for within the Standard Model (SM) of particle physics. Evidence
for dark matter is inferred from gravitational phenomena over very different
length scales, including galactic rotation curves, the dynamics and
gravitational lensing of galaxy clusters, and the anisotropies of the cosmic
microwave background~\cite{Planck:2018vyg}. Despite the robustness of
this gravitational evidence, the microscopic nature of dark matter remains
unknown.

For several decades, weakly interacting massive particles (WIMPs) provided
one of the main benchmarks for experimental dark-matter searches. Their
connection to physics at the electroweak scale and the possibility of
obtaining the observed relic abundance through thermal freeze-out motivated
an extensive experimental program based on direct detection, indirect
searches, and collider experiments~\cite{Arcadi:2017kky,Arcadi:2024ukq,Cirelli:2024ssz}. The absence
of a conclusive WIMP signal, together with increasingly stringent
experimental constraints, has motivated a substantial broadening of the
search program beyond the traditional weak-scale paradigm
~\cite{Alexander:2016aln}.

Particular attention has consequently been given to dark sectors containing
new particles at sub-GeV and MeV scales. Such sectors may contain light dark
matter states as well as new mediators connecting otherwise secluded degrees
of freedom to the SM. Two well-motivated examples
are dark photons and axion-like particles (ALPs). A dark photon, associated
with an additional Abelian gauge symmetry, can interact with electrically
charged SM particles through kinetic mixing with the ordinary photon
~\cite{Holdom:1985ag,Park:2017}. ALPs instead arise naturally as light
pseudoscalar degrees of freedom and can interact with photons, electrons,
and nucleons through a variety of effective operators
~\cite{Irastorza:2018dyq,DiLuzio:2020wdo,AristizabalSierra:2020rom}. Depending on the underlying theory, these
particles can themselves participate in the dark-matter sector or provide
light portals through which a hidden sector can be experimentally probed.

The search for light and weakly coupled states poses a distinct experimental
challenge. As the relevant masses and momentum transfers decrease, their
signals can fall below the thresholds of conventional high-energy searches,
while their small couplings imply correspondingly suppressed production and
interaction rates. These considerations motivate experiments that combine
intense particle sources, short source - detector distances, low detection
thresholds, and well-controlled backgrounds. Nuclear reactors provide an
interesting environment in this respect. In addition to their intense
antineutrino emission, reactor cores generate a large flux of photons with
energies extending into the MeV range through fission, radiative capture, and
the de-excitation of fission products and other nuclei
~\cite{Roos:1959,Bechteler:1984FRJ1,AristizabalSierra:2020rom}. These photons, as well as specific magnetic
nuclear transitions in the reactor environment, can act as sources of light
weakly coupled particles.

Reactor experiments can therefore provide laboratory probes of dark-sector
states through production and detection mechanisms that are complementary to
those explored in conventional dark matter searches. Reactor photons can
produce dark photons through Compton-like conversion on electrons, while
ALPs can be generated through Primakoff conversion, Compton-like scattering,
or magnetic nuclear de-excitation, depending on their dominant coupling
~\cite{Park:2017,Dent:2019ueq,AristizabalSierra:2020rom}. Once produced, these particles may propagate out
of the reactor core and generate observable signatures in a nearby detector
through scattering, absorption, or decay. The combination of an intense
source with a short baseline can be particularly advantageous for particles
whose interactions are too weak to be efficiently tested in smaller-scale
laboratory sources.

Reactor searches for axions emitted in magnetic nuclear transitions, motivated by the original arguments of Refs.~\cite{Weinberg:1978,DonnellyAxions:1978}, date back to the Bugey experiment~\cite{Cavaignac:1983}, which was later extended by TEXONO~\cite{TEXONO:2006axion}. More recently, the NEON experiment has set new constraints on ALPs at a nuclear reactor as is shown in ~\cite{NEON:2024kwv}. A different search was proposed in Ref.~\cite{Dent:2019ueq} based on ALP scattering at reactor neutrino experiments.  See Ref.~\cite{AristizabalSierra:2020rom} for a systematic study of ALPs at reactor experiments.

In this work we derive the projected sensitivities of CONUS+, a germanium experiment at $20.7$~m from the $3.6$~GW core of the Leibstadt reactor~\cite{ConusPlus:2024}, and of a conceivable $100$-kg CONUS100 upgrade~\cite{Conus100:2025}, taking into account their detector geometry and a per-fission normalization of the nuclear-line yields, and compare them with current experimental limits.

The paper is organized as follows. The first part is
independent of a specific detector.  In Sec.~\ref{sec:reactor_source} we
describe reactor gamma rays as a source of light weakly coupled particles,
both through the continuous photon spectrum and through discrete magnetic
nuclear transitions.  Section~\ref{darkphotonmodel} introduces the
dark photon scenario and its production and detection mechanisms, and in
Sec.~\ref{Model} we present the ALP framework and the production and
detection channels associated with photon, electron, and nucleon couplings.
In Sec.~\ref{sec:conus} we apply this framework to CONUS+ and to the CONUS100
benchmark: we describe the experimental setup and the effective detector
geometry, show representative event rates, and discuss the projected
sensitivities for the different dark photon and ALP scenarios in
Sec.~\ref{Discussion}.  Finally, Sec.~\ref{sec:conclusions} summarizes our
main results and conclusions.

\section{Reactor gamma rays as a source of light dark-sector particles}
\label{sec:reactor_source}

\subsection{Continuous reactor photon spectrum}
\label{subsec:reactor_gamma_source}

The production of weakly coupled particles in a nuclear reactor is driven by the intense prompt photon emission generated by fission fragments and their subsequent de-excitations. We model the continuous reactor photon spectrum through the empirical parameterization~\cite{Bechteler:1984FRJ1,AristizabalSierra:2020rom}, 
\begin{equation}
    \frac{d\dot N_\gamma}{dE_\gamma}
    =
    5.8\times10^{17}
    \left(\frac{P}{\mathrm{MW}}\right)
    \exp\!\left[-1.1\frac{E_\gamma}{\mathrm{MeV}}\right]
    \;\mathrm{MeV}^{-1}\mathrm{s}^{-1},
    \label{eq:reactor_gamma_spectrum}
\end{equation}
where $P$ is the reactor thermal power. Although this quantity is often
referred to as the reactor photon flux, its units show that it represents the total differential photon emission rate of the core, rather than a flux per unit area. The flux at a detector at a distance $L$ from the core is obtained by dividing the production rate of the new particle by $4\pi L^2$.

\begin{figure}[tbp]
    \centering
    \includegraphics[width=0.90\linewidth]{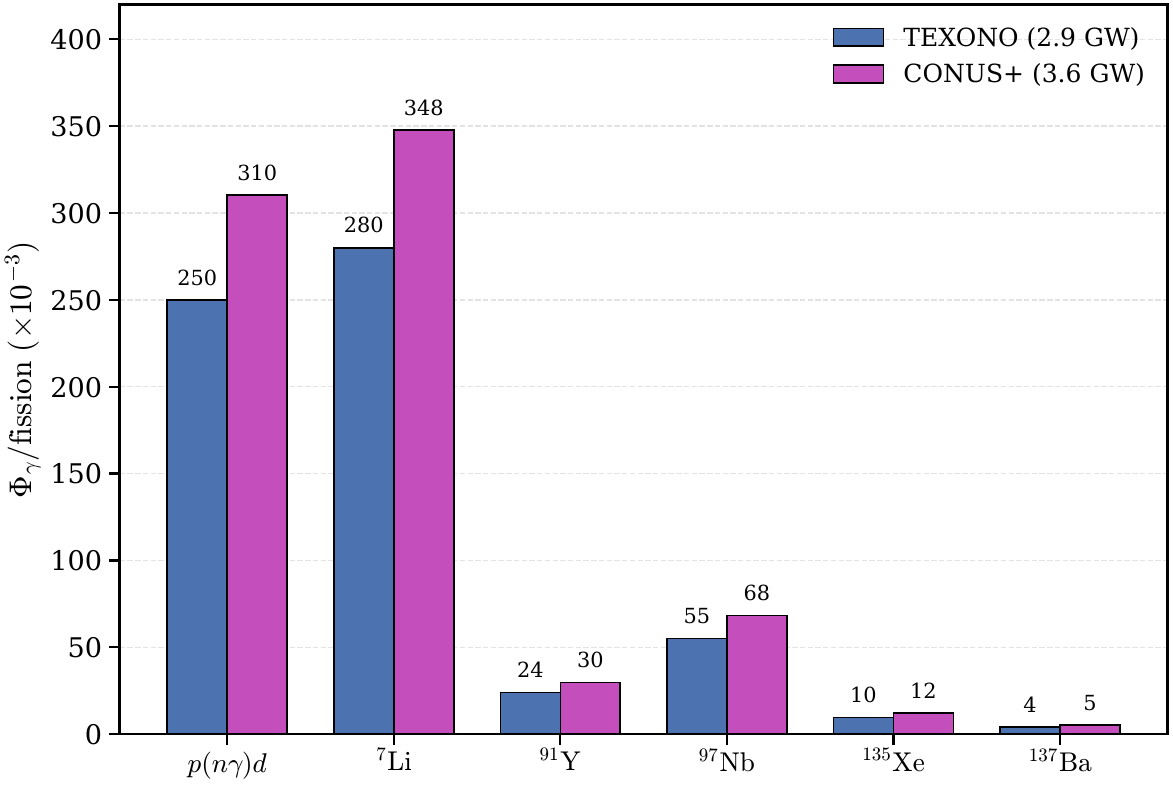}
    \caption{Photon yield per fission, $\Phi_{\gamma,i}$, of the six magnetic nuclear
    transitions, taken from
    TEXONO~\cite{TEXONO:2006axion,AristizabalSierra:2020rom} (left axis), and the corresponding
    emission rate at the $3.6~\mathrm{GW}$ KKL core, $R_f(P)\,\Phi_{\gamma,i}$, with $R_f$ from
    Eq.~\eqref{eq:gann_fission_rate} (right axis).}
    \label{fig:gann_line_yields}
\end{figure}

In our case,{Eq.~\eqref{eq:reactor_gamma_spectrum} is evaluated up to
$E_{\max}=10~\mathrm{MeV}$.  The lower limit depends on the process: for example, $25~\mathrm{keV}$ for ALP production, as in Ref.~\cite{AristizabalSierra:2020rom},
and $0.2~\mathrm{MeV}$ for dark photon production, as in Ref.~\cite{Park:2017}. The
boundaries define the theoretical integration window and should not be
identified with the instrumental energy acceptance of a specific detector.
Equation~\eqref{eq:reactor_gamma_spectrum} is an empirical parameterization, not the result of a photon transport simulation of a specific core. We assume that the photon emission scales linearly with the reactor thermal power and neglect variations associated with fuel evolution, core geometry, structural materials and reactor operation; the core is treated as a pointlike, isotropic source. For a reactor experiment such as CONUS+, at the $3.6~\mathrm{GW}$ KKL core (Table~\ref{tab:alp_benchmarks}), the continuous spectrum is simply evaluated with $P=3.6~\mathrm{GW}$.  The nuclear-line emission is instead taken from TEXONO, at a $2.9~\mathrm{GW}$ reactor~\cite{TEXONO:2006axion}, and scaled to KKL through the fission rate of Eq.~\eqref{eq:gann_fission_rate}, i.e.\ by the ratio $3.6/2.9$; the resulting rates are shown on the right axis of Fig.~\ref{fig:gann_line_yields}.

This continuous photon spectrum is used below for dark photon production and for ALP production through Primakoff conversion and Compton-like scattering.
ALP production through nuclear de-excitation is treated separately because it is determined by the discrete photon yields of specific magnetic nuclear transitions rather than by the continuous spectrum of
Eq.~\eqref{eq:reactor_gamma_spectrum}.

\subsection{Discrete magnetic nuclear transitions}
\label{par:gann_transitions}
\label{subsec:reactor_lines}

Besides the continuous spectrum, the reactor environment emits
monochromatic photons in specific nuclear transitions.  Magnetic ($MJ$)
transitions are of particular interest, because a pseudoscalar particle
coupled to nucleons can be emitted instead of the photon
(Sec.~\ref{subsec:gann}).  We adopt the six magnetic transitions used in the
TEXONO analysis~\cite{TEXONO:2006axion} and in
Ref.~\cite{AristizabalSierra:2020rom}.  Thermal-neutron capture on hydrogen in
the reactor coolant, $p(n,\gamma)d$, produces a $2.230~\mathrm{MeV}$ line that
is predominantly an isovector $M1$ transition.  The
$^{10}\mathrm{B}(n,\alpha){}^{7}\mathrm{Li}^{*}$ reaction populates the first
excited state of $^{7}\mathrm{Li}$, whose predominantly $M1$ de-excitation
produces a $478~\mathrm{keV}$ line.  The remaining four sources are magnetic
$M4$ transitions in the fission fragments $^{91}\mathrm{Y}^{*}$
($555~\mathrm{keV}$), $^{97}\mathrm{Nb}^{*}$ ($743~\mathrm{keV}$),
$^{135}\mathrm{Xe}^{*}$ ($526~\mathrm{keV}$), and
$^{137}\mathrm{Ba}^{*}$ ($662~\mathrm{keV}$).  The neutron-capture line and
the four $M4$ lines were considered in the Bugey search
~\cite{Cavaignac:1983}; TEXONO subsequently included the
$^{7}\mathrm{Li}^{*}$ transition~\cite{TEXONO:2006axion},
using nuclear-structure calculations from Refs.~\cite{Krcmar:2001,TEXONO:2006axion}.

The photon-line yields per fission adopted in the calculation are summarized
in Fig.~\ref{fig:gann_line_yields}.  We denote the yield of line $i$ by
$\Phi_{\gamma,i}$, which is dimensionless when quoted as photons per fission.
The TEXONO per-fission yields~\cite{TEXONO:2006axion,AristizabalSierra:2020rom}
are adopted as benchmark values for the reactor configurations considered
here; differences in fuel evolution, control-rod composition, neutron
transport and core geometry are neglected, so these yields are an approximate
reactor model rather than the result of a dedicated KKL core simulation.
Because it is a per-fission quantity, $\Phi_{\gamma,i}$ is not multiplied by a
reactor-power ratio.  Reactor power enters only through the fission rate
defined below.  This prescription avoids counting the power dependence twice:
since $R_f\propto P$, it is equivalent to the single rescaling by the power
ratio described in Sec.~\ref{subsec:reactor_gamma_source},
$R_f(P)\,\Phi_{\gamma,i}=(P/P_{\rm TEXONO})\,R_f(P_{\rm TEXONO})\,\Phi_{\gamma,i}$.
It differs from Ref.~\cite{AristizabalSierra:2020rom}, where the TEXONO yields
were additionally rescaled by $4/2.9$ (the ratio of reactor powers); for the
same reactor this would increase the line rates by $38\%$ and strengthen the
projected bounds on $|g_{a\gamma\gamma}g_{ann}^{(1)}|$ and
$|g_{aee}g_{ann}^{(1)}|$ by a factor $\simeq1.17$}.

The fission rate is normalized following
Ref.~\cite{AristizabalSierra:2020rom} as
\begin{equation}
    R_f(P)=1.248\times10^{20}
    \left(\frac{P}{4~\mathrm{GW}}\right)\mathrm{s}^{-1},
    \label{eq:gann_fission_rate}
\end{equation}
so that the emission rate of line $i$ is $R_f(P)\,\Phi_{\gamma,i}$.  This
simple conversion assumes that the fission rate scales linearly with thermal
power; the normalization corresponds to an energy release of
$200~\mathrm{MeV}$ per fission,
$4~\mathrm{GW}/200~\mathrm{MeV}\simeq1.248\times10^{20}~\mathrm{s^{-1}}$.

\section{Dark Photon}\label{darkphotonmodel}
\subsection{The model}
\label{subsec:dark_photon_model}

We consider a minimal dark photon scenario in which the SM
electromagnetic sector is supplemented by an additional Abelian gauge field
$A'_\mu$ associated with a hidden $U(1)_D$ symmetry. At low energies, the
visible and dark gauge sectors can communicate through kinetic mixing,
\begin{equation}
    \mathcal{L}
    =
    -\frac{1}{4}F_{\mu\nu}F^{\mu\nu}
    -\frac{1}{4}F'_{\mu\nu}F'^{\mu\nu}
    -\frac{\epsilon}{2}F_{\mu\nu}F'^{\mu\nu}
    +\frac{1}{2}m_{A'}^{2}A'_\mu A'^{\mu}
    -e A_\mu J_{\rm EM}^{\mu},
    \label{eq:dark_photon_kinetic_mixing}
\end{equation}
where $F_{\mu\nu}$ and $F'_{\mu\nu}$ are the electromagnetic and
dark photon field strength tensors, respectively, $m_{A'}$ denotes the
dark photon mass, and $\epsilon$ is the kinetic-mixing parameter.

After diagonalizing and canonically normalizing the gauge kinetic terms, the
dark photon acquires an effective coupling to the electromagnetic current.
To leading order in $\epsilon$, the interaction relevant for the present
analysis can be written as
\begin{equation}
    \mathcal{L}_{\rm int}
    =
    -\epsilon e A'_\mu J_{\rm EM}^{\mu}.
    \label{eq:dark_photon_effective_interaction}
\end{equation}
In particular, the coupling to electrons is
\begin{equation}
    \mathcal{L}_{A'e}
    =
    -\epsilon e
    A'_\mu \bar e\gamma^\mu e .
    \label{eq:dark_photon_electron_coupling}
\end{equation}
Therefore, the phenomenology considered here is fully specified by the two
independent parameters: $m_{A'}$ and $\epsilon$.

\begin{figure}[H]
    \centering

    \begin{subfigure}[b]{0.45\textwidth}
        \centering
        \begin{tikzpicture}[scale=0.55, every node/.style={scale=0.85}]
        \begin{feynman}
        \vertex (v1);
        \vertex [right=2cm of v1, dot] (v2) {};
        \vertex [above left=2.3cm of v1] (g_in) {\(\gamma\)};
        \vertex [below left=2cm of v1] (e_in) {\(e\)};
        \vertex [above right=2.3cm of v2] (Ap_out) {\(A'\)};
        \vertex [below right=2cm of v2] (e_out) {\(e\)};
        \diagram* {
          (g_in) -- [photon] (v1),
          (e_in) -- [plain] (v1),
          (v1) -- [plain, edge label'=\(e\)] (v2),
          (v2) -- [photon] (Ap_out),
          (v2) -- [plain] (e_out),
        };
        \end{feynman}
        \end{tikzpicture}
        \caption{Compton-like scattering}
        \label{fig:darkphoton_production}
    \end{subfigure}
    \hfill
    \begin{subfigure}[b]{0.45\textwidth}
        \centering
        \begin{tikzpicture}[scale=0.55, every node/.style={scale=0.85}]
        \begin{feynman}
        \vertex (v1);
        \vertex [right=2cm of v1, dot] (v2) {};
        \vertex [above left=2.2cm of v1] (Ap_in) {\(A'\)};
        \vertex [below left=2cm of v1] (e_in) {\(e\)};
        \vertex [above right=2.3cm of v2] (g_out) {\(\gamma\)};
        \vertex [below right=2cm of v2] (e_out) {\(e\)};
        \diagram* {
          (Ap_in) -- [photon] (v1),
          (e_in) -- [plain] (v1),
          (v1) -- [plain, edge label'=\(e\)] (v2),
          (v2) -- [photon] (g_out),
          (v2) -- [plain] (e_out),
        };
        \end{feynman}
        \end{tikzpicture}
        \caption{Inverse Compton-like scattering}
        \label{fig:darkphoton_detection}
    \end{subfigure}

    \caption{Dark photon production and detection processes. (a) Production via Compton-like scattering $\gamma + e^- \to A' + e^-$. (b) Detection via the inverse process, $A' + e^- \to \gamma + e^-$.}
    \label{fig:darkphoton_processes}
\end{figure}
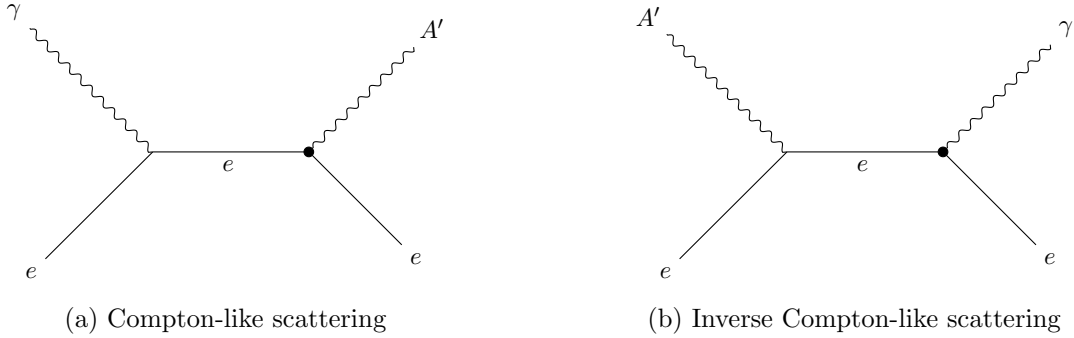

The interaction in Eq.~\eqref{eq:dark_photon_electron_coupling} allows
reactor photons to produce dark photons through Compton-like conversion on
electrons, while dark photons reaching the germanium detector can be observed through the crossed process (Fig.~\ref{fig:darkphoton_processes}).
Since the same effective interaction controls production and detection, both cross sections are proportional to $\epsilon^2$, and the event yield scales approximately as $\epsilon^4$ in the absence of propagation effects.

In this work, we restrict the analysis to $m_{A'}<2m_e$, so that the tree-level decay $A'\rightarrow e^+e^-$ is kinematically forbidden. We also consider only $m_{A'}\geq100~\mathrm{eV}$: in a medium the photon acquires an effective mass equal to the plasma frequency, $m_\gamma\simeq20~\mathrm{eV}$ in the reactor and $\simeq44~\mathrm{eV}$ in germanium, and for $m_{A'}\lesssim m_\gamma$ both the production and the detection probabilities are suppressed by $(m_{A'}/m_\gamma)^4$~\cite{Danilov:2018tqg}, while for $m_{A'}\gtrsim100~\mathrm{eV}$ this effect changes $\epsilon$ by less than $\sim15\%$. In the minimal visible-sector scenario considered here, the remaining loop induced visible decays are sufficiently suppressed over the parameter region of interest and are neglected. We also assume that no lighter dark-sector states are available as invisible decay products. The production and detection rates associated with these processes are discussed in the following subsection.

\subsection{Production and detection mechanism}

\subsubsection{Production}
The production channel under consideration is the inelastic scattering of a reactor photon off an atomic electron, $\gamma + e^- \to A' + e^-$ (see \autoref{fig:darkphoton_production}), where $A'$ denotes the massive dark photon. We use the exact tree-level cross section for a massive vector, in which the dark photon mass is kept in the matrix element, in the kinematics and in the production threshold $\omega\geq m_{A'}+m_{A'}^2/(2m_e)$, without assuming $m_{A'}\ll\omega$. For a dark photon mass $m_{A'} \ll m_e$ (where $m_e = 0.511$ MeV), it reduces to the ordinary Klein-Nishina differential cross-section evaluated at the same outgoing energy:

\begin{equation}
\frac{d\sigma_{\gamma \to A'}}{dE_{A'}}(\omega, E_{A'}) \;\approx\; \epsilon^2 \,
\frac{d\sigma_{\rm KN}}{d\omega'}(\omega, \omega')\bigg|_{\omega' = E_{A'}},
\label{eq:prod_xsec}
\end{equation}

where $E_{A'}$ is the total energy of the produced dark photon. Ref.~\cite{Park:2017} replaced the total photon cross section of the reactor medium by the Klein-Nishina one.  Below $\sim1~\mathrm{MeV}$, however, photoelectric absorption is important in uranium, and the Compton fraction of the total cross section is only $11$--$42\%$ between $0.25$ and $0.6~\mathrm{MeV}$.  We therefore use the total photon cross section of uranium from XCOM~\cite{BergerXCOM}, $\sigma_{\rm tot}^{U}(\omega)$, and multiply the cross section per electron of Eq.~\eqref{eq:prod_xsec} by $Z_U=92$.

The differential production rate of dark photons in the reactor core is obtained by folding the conversion probability with the source spectrum:

\begin{equation}
\mathcal{F}_{A'}(E_{A'}) \equiv \frac{d\dot N_{A'}}{dE_{A'}}
= \int_{\omega_{\rm min}}^{\omega_{\rm max}} d\omega\;
\frac{Z_U}{\sigma_{\rm tot}^{U}(\omega)}
\frac{d\sigma_{\gamma \to A'}}{dE_{A'}}(\omega, E_{A'})
\frac{d\dot N_{\gamma}}{d\omega},
\label{eq:production_conv}
\end{equation}

where $d\dot N_\gamma/d\omega$ is the photon emission rate of Eq.~\eqref{eq:reactor_gamma_spectrum} and the integration limits are implicitly constrained by the kinematic requirement that the outgoing dark photon energy must lie within the physical Compton continuum for the given incident photon energy. Equation~\eqref{eq:production_conv} is evaluated numerically with incident photon energies spanning $\omega \in [0.2, 10]$~MeV, the lower edge being the one adopted for the parameterization of Eq.~\eqref{eq:reactor_gamma_spectrum} in Ref.~\cite{Park:2017}.

\begin{figure}[h!]
    \centering
    \includegraphics[width=1\linewidth]{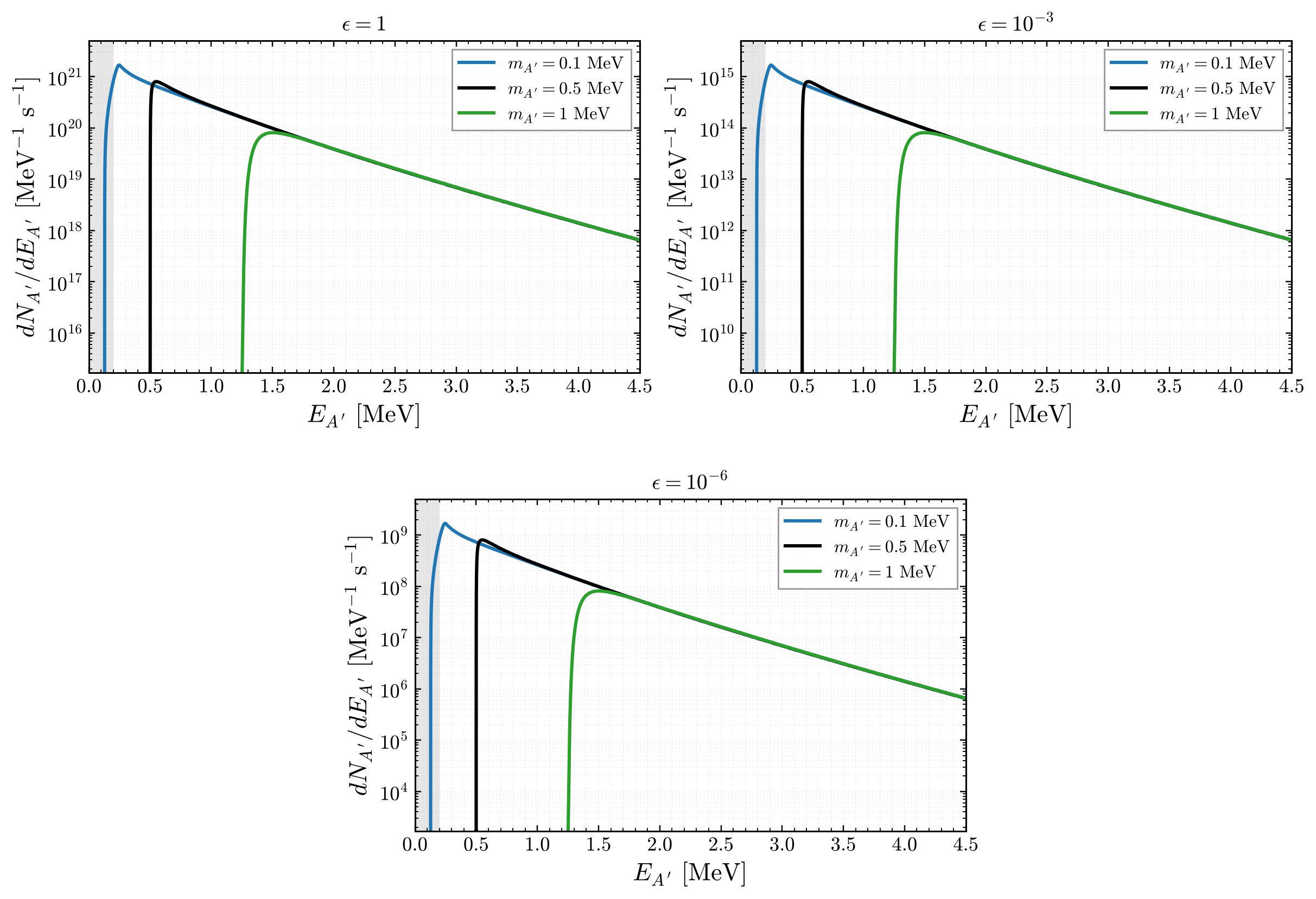}
    \caption{dark photon production spectrum at the KKL core ($3.6~\mathrm{GW}$) for
    $\epsilon=1$ (top left), $10^{-3}$ (top right) and $10^{-6}$ (bottom), and $m_{A'}=0.1$ (blue), $0.5$ (black) and $1~\mathrm{MeV}$ (green), from reactor photons above $0.2~\mathrm{MeV}$ (the shaded area is below this energy).}
    \label{fig:DP_production_flux}
\end{figure}

Two refinements of this treatment have been discussed in the literature.  In the oscillation description of Ref.~\cite{Danilov:2018tqg}, valid for $m_\gamma\ll m_{A'}\ll\omega$, every reactor photon converts with probability $\epsilon^2$ independently of the absorption in the medium, and only the two transverse polarizations are produced.  In addition, Compton-scattered photons produce further dark photons~\cite{Du:2023tqs}; Ref.~\cite{Demidov:2026cmt} finds this to increase the flux by a factor $1.2$--$2.6$ between $2$ and $0.2~\mathrm{MeV}$. For our CONUS+ and CONUS100 benchmarks these treatments would lower the projected $\epsilon$ by about $30$--$35\%$ for $m_{A'}\ll\omega$, where they apply.  The calculation adopted here, which keeps the full mass dependence needed for $m_{A'}\sim\omega$, is therefore conservative.

\subsubsection{Detection}
Once produced, the dark photons propagate to the detector with negligible attenuation, as the decay length far exceeds reactor--detector baselines of a few tens of metres. Detection proceeds via the time-reversed process, $A' + e^- \to \gamma + e^-$ (see \autoref{fig:darkphoton_detection}). The integrated detection cross-section for a given dark photon energy $E_{A'}$ is

\begin{equation}
\Sigma_{A'}(E_{A'})\xrightarrow{m_{A'}\to0}\frac{2}{3}\,\epsilon^2
\int_{E_{\min}}^{E_{\max}} dE_R \,
\frac{d\sigma_{\rm KN}}{dE_R}(\omega=E_{A'}, E_R),
\label{eq:detection_xsec}
\end{equation}

where $E_R$ denotes the energy of the recoil photon produced in the final state. The factor $2/3$ originates from the averaging over the three helicity/polarisation states of the massive dark photon. For finite $m_{A'}$ we obtain $\Sigma_{A'}$ from the exact production cross section by detailed balance, which keeps the $2/3$ factor and the ratio of the phase-space momenta.  For $m_{A'}\ll E_{A'}$, where only transverse dark photons are produced~\cite{Danilov:2018tqg}, the $2/3$ factor makes the detection rate conservative by a factor $3/2$.  In the massless limit the kinematic boundaries of the integral are determined by the Compton edge:

\begin{equation}
E_{\min}(E_{A'}) = \frac{E_{A'}}{1 + 2E_{A'}/m_e}, \qquad
E_{\max}(E_{A'}) = E_{A'}.
\label{eq:recoil_limits}
\end{equation}

For the purposes of computing the differential event rate, the integration is performed over the entire kinematically accessible range, corresponding to a detector with full acceptance and zero energy threshold. Assuming that the final-state photon and electron are fully absorbed, the deposited energy is $E_{\rm dep}=E_{A'}$, and the number of events in an accepted window $[E_{\rm dep}^{\rm min},E_{\rm dep}^{\rm max}]$ is
\begin{equation}
    N_{\rm sig}
    =
    \frac{N_e\,\Delta t}{4\pi L^2}
    \int_{\max(E_{\rm dep}^{\rm min},\,m_{A'})}^{E_{\rm dep}^{\rm max}}
    dE_{A'}\,
    \mathcal{F}_{A'}(E_{A'})\,\Sigma_{A'}(E_{A'}),
    \label{eq:dp_signal_events}
\end{equation}
where $N_e$ is the number of electrons in the germanium target, $\Delta t$ the exposure and $L$ the reactor--detector distance.  The lower limit $m_{A'}$ is reached only for $m_{A'}<m_e$: for heavier dark photons the kinematics of $\gamma e^-\to A'e^-$ on an electron at rest requires $E_{A'}\geq(m_{A'}^2+m_e^2)/(2m_e)$, e.g.\ $1.23~\mathrm{MeV}$ for $m_{A'}=1~\mathrm{MeV}$, so the spectra of Figs.~\ref{fig:DP_production_flux} and~\ref{fig:DP_rates_grid} start above $E_{A'}=m_{A'}$.  Since production and detection each contribute $\epsilon^2$,
\begin{equation}
    N_{\rm sig}=\epsilon^4\,N_0(m_{A'}),
    \label{eq:dp_epsilon_scaling}
\end{equation}
where $N_0(m_{A'})$ is the yield evaluated at $\epsilon=1$.

\section{Axion-like Particles}
\label{Model}

\subsection{Model}
\label{subsec:alp_model}

We consider a generic axion-like particle (ALP), denoted by $a$, coupled to
photons, electrons, and nucleons.  At the energies relevant for reactor
experiments, the interactions are parametrized by the effective Lagrangian
~\cite{AristizabalSierra:2020rom,Irastorza:2018dyq}
\begin{equation}
\label{eq:interactions-ALP}
\mathcal{L}
=
\mathcal{L}_a
-\frac{1}{4}\,g_{a\gamma\gamma}\,
a F_{\mu\nu}\widetilde{F}^{\mu\nu}
-i g_{aee}\,a\,\bar e\gamma_5 e
-i a\,\bar N\gamma_5
\left(
    g_{ann}^{(0)}+\tau_3 g_{ann}^{(1)}
\right)N ,
\end{equation}
where
\begin{equation}
    \mathcal{L}_a
    =
    \frac{1}{2}\partial_\mu a\,\partial^\mu a
    -\frac{1}{2}m_a^2a^2
\end{equation}
contains the free ALP kinetic and mass terms.  Here $F_{\mu\nu}$ is the
electromagnetic field-strength tensor,
$\widetilde{F}^{\mu\nu}=\epsilon^{\mu\nu\rho\sigma}F_{\rho\sigma}/2$ is its
dual, $N=(p,n)^T$ denotes the proton--neutron isospin doublet, and $\tau_3$
is the third Pauli matrix in isospin space.  The couplings
$g_{ann}^{(0)}$ and $g_{ann}^{(1)}$ are the isoscalar and isovector
ALP--nucleon couplings, respectively.

The photon coupling $g_{a\gamma\gamma}$ has dimensions of inverse energy,
whereas the pseudoscalar electron and nucleon couplings in
Eq.~\eqref{eq:interactions-ALP} are dimensionless.  The fermionic terms can
be related to the derivative ALP--fermion interaction
~\cite{Irastorza:2018dyq,DiLuzio:2020wdo,AristizabalSierra:2020rom}
\begin{equation}
    \mathcal{L}_{af}
    =
    \frac{C_{af}}{2f_a}
    \bar f\gamma^\mu\gamma_5 f\,
    \partial_\mu a,
    \qquad
    f=e,p,n,
    \label{eq:alp_derivative_coupling}
\end{equation}
which, using the fermion equations of motion, gives
\begin{equation}
    g_{aff}
    =
    \frac{m_f C_{af}}{f_a}.
    \label{eq:alp_fermion_relation}
\end{equation}
Similarly, the photon coupling can be written schematically as
\begin{equation}
    g_{a\gamma\gamma}
    =
    \frac{\alpha}{2\pi}
    \frac{C_{a\gamma}}{f_a},
    \label{eq:alp_photon_relation}
\end{equation}
where $m_f$ is the fermion mass, $\alpha$ is the electromagnetic
fine-structure constant, $f_a$ denotes the characteristic ALP interaction
scale, and $C_{af}$ and $C_{a\gamma}$ are model-dependent dimensionless
coefficients.

We adopt a phenomenological ALP framework rather than a specific ultraviolet
realization.  The ALP mass and the effective couplings are therefore treated
as independent parameters, and no QCD-axion mass--coupling relation is
imposed.  This allows the photon-, electron-, and nucleon-mediated reactor
signatures to be analyzed separately before considering the mixed-coupling
cases associated with nuclear production.

\subsection{Production and detection mechanisms}

\subsubsection{Photon--ALP coupling}
\label{subsec:gagg}

We first consider the single-coupling scenario in which
$g_{a\gamma\gamma}$ is non-zero, while the electron and nucleon couplings in
Eq.~\eqref{eq:interactions-ALP} are set to zero.  Reactor photons then produce
ALPs through Primakoff conversion, $\gamma+N\to a+N$, while detection in
germanium proceeds through inverse Primakoff scattering,
$a+N\to\gamma+N$, or through $a\to\gamma\gamma$ decays occurring inside the effective detector volume.  Loop induced interactions and decay modes not associated with $g_{a\gamma\gamma}$ are neglected in this single coupling benchmark.  The calculation follows the reactor treatment of Ref.~\cite{AristizabalSierra:2020rom}, with the reactor source of Sec.~\ref{sec:reactor_source}; the experimental benchmarks are specified in Sec.~\ref{sec:conus}.

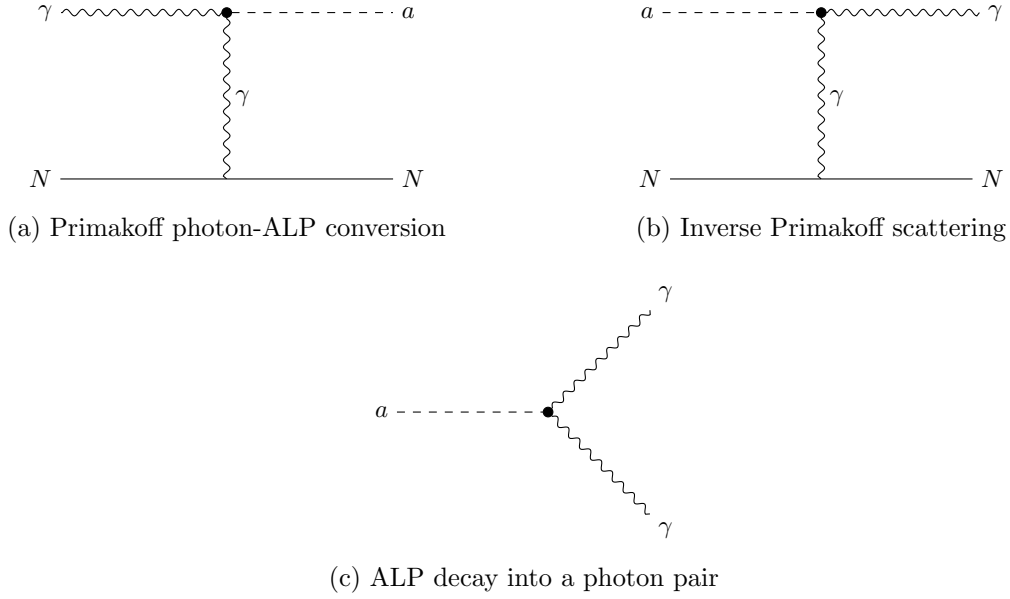
\begin{figure}
    \centering

    \begin{subfigure}[b]{0.48\textwidth}
        \centering
        \begin{tikzpicture}[scale=0.9, every node/.style={scale=0.85}]
        \begin{feynman}
        \vertex (g_in) {\(\gamma\)};
        \vertex [right=2.4 of g_in, dot] (v1) {};
        \vertex [right=2.4 of v1] (a_out) {\(a\)};
        \vertex [below=2.2 of v1] (v2);
        \vertex [left=2.2 of v2] (N_in) {\(N\)};
        \vertex [right=2.2 of v2] (N_out) {\(N\)};
        \diagram* {
          (g_in) -- [photon] (v1) -- [scalar] (a_out),
          (v1) -- [photon, edge label=\(\gamma\)] (v2),
          (N_in) -- [plain] (v2) -- [plain] (N_out),
        };
        \end{feynman}
        \end{tikzpicture}
        \caption{Primakoff photon-ALP conversion}
        \label{fig:primakoff_process}
    \end{subfigure}
    \hfill
    \begin{subfigure}[b]{0.48\textwidth}
        \centering
        \begin{tikzpicture}[scale=0.9, every node/.style={scale=0.85}]
        \begin{feynman}
        \vertex (a_in) {\(a\)};
        \vertex [right=2.3 of a_in, dot] (v1) {};
        \vertex [right=2.3 of v1] (g_out) {\(\gamma\)};
        \vertex [below=2.2 of v1] (v2);
        \vertex [left=2 of v2] (N_in) {\(N\)};
        \vertex [right=2 of v2] (N_out) {\(N\)};
        \diagram* {
          (a_in) -- [scalar] (v1) -- [photon] (g_out),
          (v1) -- [photon, edge label=\(\gamma\)] (v2),
          (N_in) -- [plain] (v2) -- [plain] (N_out),
        };
        \end{feynman}
        \end{tikzpicture}
        \caption{Inverse Primakoff scattering}
        \label{fig:inverse_primakoff}
    \end{subfigure}

    \vspace{1em}

    \begin{subfigure}[b]{0.48\textwidth}
        \centering
        \begin{tikzpicture}[scale=0.9, every node/.style={scale=0.85}]
        \begin{feynman}
        \vertex (a_in) {\(a\)};
        \vertex [right=2.2 of a_in, dot] (v1) {};
        \vertex [above right=2.2 of v1] (g_out1) {\(\gamma\)};
        \vertex [below right=2.2 of v1] (g_out2) {\(\gamma\)};
        \diagram* {
          (a_in) -- [scalar] (v1),
          (v1) -- [photon] (g_out1),
          (v1) -- [photon] (g_out2),
        };
        \end{feynman}
        \end{tikzpicture}
        \caption{ALP decay into a photon pair}
        \label{fig:alp_decay}
    \end{subfigure}

    \caption{ALP production and detection processes through the photon-ALP coupling $g_{a\gamma\gamma}$. (a) ALP production via Primakoff conversion $\gamma + N \to a + N$. (b) and (c) are the detection mechanisms considered: (b) inverse Primakoff scattering $a + N \to \gamma + N$ and (c) ALP decay into a photon pair, $a \to \gamma\gamma$.}
    \label{fig:primakoff_processes}
\end{figure}

\paragraph{Reactor production and propagation}
\label{subsec:gagg_production_propagation}

We use the reactor photon source of Sec.~\ref{subsec:reactor_gamma_source} and, as in Ref.~\cite{AristizabalSierra:2020rom}, an effective $^{235}\mathrm{U}$ production medium with $Z_U=92$, $A_U=235.04$ and $M_U=A_U\times931.5~\mathrm{MeV}$.  To distinguish the incident reactor photon from photons produced in the detector, we denote its energy by $E'_\gamma$.

For a target nucleus with mass $M_N$ and atomic number $Z$, the differential
Primakoff production cross section is taken from the exact expression of
Refs.~\cite{Aloni:2019ruo,AristizabalSierra:2020rom}
\begin{multline}
    \frac{d\sigma_{P}^{P}}{dt}
    =
    2\alpha Z^2F^2(t)g_{a\gamma\gamma}^{2}
    \frac{M_N^4}
    {t^2(M_N^2-s)^2(t-4M_N^2)^2}
    \\
    \times
    \left\{
    m_a^2t(M_N^2+s)-m_a^4M_N^2
    -t\left[(M_N^2-s)^2+st\right]
    \right\},
    \label{eq:primakoff_dsigma_gagg}
\end{multline}
where
\begin{equation}
    s=M_N^2+2M_NE'_\gamma
\end{equation}
and $t<0$ is the Mandelstam variable associated with the transfer of momentum.
Equation~\eqref{eq:primakoff_dsigma_gagg} does not rely on either the
forward-scattering or the massless-ALP approximation. The total production
cross section is obtained by integrating over the kinematically allowed
momentum transfer,
\begin{equation}
    \sigma_{P}^{P}(E'_\gamma;N)
    =
    \int_{t_{\min}}^{t_{\max}}
    \frac{d\sigma_{P}^{P}}{dt}\,dt\,,
    \label{eq:primakoff_sigma_gagg}
\end{equation}
with the exact two-body limits used in
Refs.~\cite{PDG:2018,AristizabalSierra:2020rom},
\begin{align}
    t_{\min}
    &=
    \frac{m_a^4}{4s}-(p_{\rm cm}+k_{\rm cm})^2\,,
    &
    t_{\max}
    &=
    \frac{m_a^4}{4s}-(p_{\rm cm}-k_{\rm cm})^2\,,
    \label{eq:t_limits_gagg}
    \\
    p_{\rm cm}
    &=
    \frac{s-M_N^2}{2\sqrt{s}}\,,
    &
    k_{\rm cm}
    &=
    \left[
    \frac{(s+m_a^2-M_N^2)^2}{4s}-m_a^2
    \right]^{1/2}.
    \label{eq:cm_momenta_gagg}
\end{align}
The reaction is kinematically allowed only when
$s\geq(M_N+m_a)^2$, which imposes the incident-photon threshold
\begin{equation}
    E'_{\gamma,\mathrm{thr}}
    =
    m_a+\frac{m_a^2}{2M_N}\,.
    \label{eq:primakoff_threshold_gagg}
\end{equation}
Together with the lower boundary of the reactor photon spectrum, this
defines the integration limit
\begin{equation}
    E_{\rm lo}(m_a)
    =
    \max\!\left(E_{\min},E'_{\gamma,\mathrm{thr}}\right).
    \label{eq:primakoff_energy_lower_gagg}
\end{equation}

Atomic screening and the loss of nuclear coherence at finite momentum transfer
are incorporated through a piecewise form factor. Defining
$q\equiv\sqrt{-t}$, for $-t\leq7.39m_e^2$, where the momentum transfer is small enough for the
atomic electrons to screen the nuclear charge~\cite{Tsai:1986tx}, we use the
atomic form factor of the Thomas--Fermi atom in the Moli\`ere
representation~\cite{Tsai:1974fa}, written for $t<0$ as in
Ref.~\cite{AristizabalSierra:2020rom},
\begin{equation}
    F_{\rm A}^2(t)
    =
    \left(
    \frac{a_0^2t}{1-a_0^2t}
    \right)^2,
    \qquad
    a_0=
    \frac{184.15}{\exp(1/2)\,Z^{1/3}m_e}\,.
    \label{eq:atomic_form_factor_gagg}
\end{equation}
For larger momentum transfer, it is replaced by the Helm nuclear form
factor~\cite{Helm:1956zz}, as in Ref.~\cite{AristizabalSierra:2020rom}
(Ref.~\cite{Aloni:2019ruo} uses instead a two-parameter Fermi form factor;
the difference is negligible here~\cite{AristizabalSierra:2019zmy}),
\begin{equation}
    F_{\rm H}^2(q)
    =
    \left[
    \frac{3j_1(qR_1)}{qR_1}
    \right]^2
    e^{-(q s_H)^2},
    \qquad
    R_1=\sqrt{R^2-5s_H^2},
    \quad
    R=1.2A^{1/3}\,\mathrm{fm},
    \quad
    s_H=0.9\,\mathrm{fm},
    \label{eq:helm_form_factor_gagg}
\end{equation}
where $j_1$ denotes the spherical Bessel function of the first kind and
$A$ is the nuclear mass number. The atomic-screening correction is particularly important in the low-mass region,
$m_a\lesssim10^4~\mathrm{eV}$, where Primakoff production is dominated by
small momentum transfers. The Helm correction has a smaller numerical effect
over most of the parameter space considered here.

Having specified the Primakoff production cross section, the corresponding
photon-to-ALP conversion probability must be normalized to all competing
photon interactions in the reactor fuel. We obtain the Standard Model photon cross section from the ``total with coherent scattering'' mass attenuation coefficient for uranium tabulated in the XCOM database
~\cite{BergerXCOM}. This quantity includes coherent and incoherent scattering, photoelectric absorption, and electron-positron pair production whenever kinematically allowed.

The XCOM mass attenuation coefficient is converted from
$\mathrm{cm^2\,g^{-1}}$ to a cross section per uranium atom according to
\begin{equation}
    \sigma_{\rm SM}^{U}(E)
    =
    \frac{m_{\rm mol}^{U}}{N_A}
    \left(\frac{\mu}{\rho}\right)_{\!\mathrm{XCOM}}(E),
    \label{eq:xcom_conversion_gagg}
\end{equation}
where $m_{\rm mol}^{U}$ is the molar mass of natural uranium and $N_A$ is Avogadro's constant.

The total interaction cross section entering the conversion probability is
therefore
\begin{equation}
    \sigma_{\rm Tot}^{U}(E)
    =
    \sigma_{\rm SM}^{U}(E)
    +
    \sigma_{P}^{P}(E;U).
    \label{eq:total_cross_section_gagg}
\end{equation}
The probability for a photon of energy $E$ to produce an ALP is consequently
$\sigma_{P}^{P}/\sigma_{\rm Tot}^{U}$. Retaining the Primakoff contribution
in the denominator guarantees that this probability remains smaller than
unity, although in the coupling region relevant for the projected sensitivity
one typically has $\sigma_{P}^{P}\ll\sigma_{\rm SM}^{U}$.

For a fixed incident-photon energy, two-body kinematics restricts the outgoing
ALP energy to a narrow interval around $E_a\simeq E'_\gamma$. Following
Ref.~\cite{AristizabalSierra:2020rom}, we therefore employ the
narrow-energy-transfer approximation
\begin{equation}
    \frac{d\sigma_{P}^{P}}{dE_a}
    \simeq
    \sigma_{P}^{P}(E_a;U)\,
    \delta(E'_\gamma-E_a).
    \label{eq:narrow_transfer_gagg}
\end{equation}
After performing the convolution with the reactor photon spectrum, the
survival-weighted differential ALP rate becomes
\begin{equation}
    \frac{d\dot N_a^{P}}{dE_a}
    =
    P_{\rm surv}(E_a)
    \frac{\sigma_{P}^{P}(E_a;U)}
    {\sigma_{\rm SM}^{U}(E_a)+\sigma_{P}^{P}(E_a;U)}
    \left.
    \frac{d\dot N_\gamma}{dE'_\gamma}
    \right|_{E'_\gamma=E_a}.
    \label{eq:alp_source_rate_gagg}
\end{equation}
This quantity has units of
$\mathrm{MeV^{-1}\,s^{-1}}$ and does not yet include geometrical dilution.
Accordingly, the differential ALP flux incident on a detector at a distance
$L$ is
\begin{equation}
    \frac{d\Phi_a^{P}}{dE_a}
    =
    \frac{1}{4\pi L^2}
    \frac{d\dot N_a^{P}}{dE_a}.
    \label{eq:alp_flux_detector_gagg}
\end{equation}

Within the single-coupling scenario considered here, the ALP decay width is
dominated by the diphoton channel~\cite{Irastorza:2018dyq,AristizabalSierra:2020rom},
\begin{equation}
    \Gamma_a
    =
    \Gamma_{a\to\gamma\gamma}
    =
    \frac{g_{a\gamma\gamma}^{2}m_a^3}{64\pi}\textcolor{red}{,}
    \label{eq:gamma_agg}
\end{equation}
where $m_a$ is the ALP mass.
For an ALP with energy $E_a$ and momentum
$p_a=\sqrt{E_a^2-m_a^2}$, the probability of surviving the propagation
distance $L$ is~\cite{Dent:2019ueq,AristizabalSierra:2020rom}
\begin{equation}
    P_{\rm surv}(E_a)
    =
    \exp\!\left[
    -\frac{L m_a\Gamma_a}{p_a\hbar c}
    \right]\textcolor{red}{,}
    \label{eq:survival_gagg}
\end{equation}
where $p_a\hbar c/(m_a\Gamma_a)$ is the ALP decay length in the laboratory frame.
The conditional probability that an ALP reaching the detector decays over an
effective mean path length $\ell_{\rm eff}$ inside the detector (Sec.~\ref{subsec:conus_geometry})is
\begin{equation}
    P_{\rm dec}(E_a)
    =
    1-
    \exp\!\left[
    -\frac{\ell_{\rm eff}m_a\Gamma_a}{p_a\hbar c}
    \right].
    \label{eq:decay_probability_gagg}
\end{equation}

\paragraph{Detection mechanisms and event yields}
\label{subsec:gagg_detection}

The ALP signal is the sum of two physically different contributions:
\begin{enumerate}
\item \textbf{Inverse Primakoff scattering
($a+\mathrm{Ge}\to\gamma+\mathrm{Ge}$).}
For inverse Primakoff detection we adopt the prescription of
Ref.~\cite{AristizabalSierra:2020rom},
\begin{equation}
\sigma_{D}^{P}(E_a;\mathrm{Ge})
=2\,\sigma_{P}^{P}(E_a;\mathrm{Ge})\,.
\label{eq:inverse_primakoff_sigma_gagg}
\end{equation}
The physical origin of the factor of two is the interchange of the photon
between the initial and final states. By microscopic reversibility the same
matrix element governs the direct and inverse processes; in the small-recoil
kinematics $E_a\simeq E_\gamma$ used here, production averages over the two
initial photon polarizations whereas the inverse process sums over the two
final photon polarizations. We therefore use Eq.~\eqref{eq:inverse_primakoff_sigma_gagg}
as the reference prescription rather than as an independent exact
detailed-balance calculation away from this kinematic limit. Here
Eq.~\eqref{eq:primakoff_dsigma_gagg} is evaluated with
$Z_{\rm Ge}=32$, $A_{\rm Ge}=72.63$, and
$M_{\rm Ge}=A_{\rm Ge}\times931.5~\mathrm{MeV}$.  The expected number of
events is
\begin{equation}
N_{\rm IP}
=\frac{M_{\rm det}N_T\Delta t}{4\pi L^2}
\int_{E_{\rm lo}(m_a)}^{E_{\max}}
\frac{d\dot N_a^{P}}{dE_a}\,
\sigma_{D}^{P}(E_a;\mathrm{Ge})\,dE_a,
\label{eq:inverse_primakoff_yield_gagg}
\end{equation}
with
$N_T=N_A/(A_{\rm Ge}\times10^{-3}~\mathrm{kg\,mol^{-1}})$ the number of
germanium nuclei per kilogram.  Nuclear recoil is neglected in the
reconstructed energy, so the outgoing photon is taken to carry essentially
the full ALP energy.

\item \textbf{Decay inside the detector ($a\to\gamma\gamma$).}
An ALP intersecting the detector can instead decay into two photons. The
corresponding event yield is
\begin{equation}
    N_{\gamma\gamma}
    =\frac{A_{\rm proj}\Delta t}{4\pi L^2}
    \int_{E_{\rm lo}(m_a)}^{E_{\max}}
    \frac{d\dot N_a^{P}}{dE_a}\,
    P_{\rm dec}(E_a)\,dE_a,
    \label{eq:diphoton_yield_gagg}
\end{equation}
where $A_{\rm proj}$ is an effective projected detector area and
$P_{\rm dec}$ is evaluated with the effective mean chord
$\ell_{\rm eff}$, with $A_{\rm proj}\ell_{\rm eff}$ equal to the detector
volume. This generalizes the $A=L_{\rm det}^2$,
$L_{\rm det}$ cube prescription of
Ref.~\cite{AristizabalSierra:2020rom}.

The effective geometry does not constitute a detector-response simulation; photon escape, annihilation photons, dead layers, inter-crystal gaps, passive material and event-topology selections are not modelled explicitly.

\end{enumerate}

The predicted photon-coupling signal is
\begin{equation}
S_{a\gamma\gamma}=N_{\rm IP}+N_{\gamma\gamma}.
\label{eq:total_signal_gagg}
\end{equation}
The effective geometries and the experimental parameters of CONUS+ and
CONUS100 are given in Sec.~\ref{sec:conus}; the detector efficiency and
background assumptions are introduced only when the projected sensitivities are constructed in Sec.~\ref{subsec:alp_discussion}.

\subsubsection{Electron--ALP coupling}
\label{subsec:gaee}

We next consider the complementary single-coupling scenario in which
$g_{aee}$ is non-zero, while $g_{a\gamma\gamma}$ and
$g_{ann}^{(0,1)}$ in Eq.~\eqref{eq:interactions-ALP} are set to zero.
Reactor photons produce ALPs through Compton-like scattering,
$\gamma e^-\to a e^-$, while detection in germanium proceeds through inverse
Compton-like scattering, axio-electric absorption, or, when kinematically
open, the decay $a\to e^+e^-$.  These are the three channels considered here, following Ref.~\cite{AristizabalSierra:2020rom}.  ALP-induced pair
production in the nuclear or electron field is not included, consistently
with the same reference treatment, where its expected yield is subleading.

\begin{figure}[H]
    \centering
    \begin{subfigure}[b]{0.45\textwidth}
        \centering
        \begin{tikzpicture}[scale=0.55, every node/.style={scale=0.85}]
        \begin{feynman}
        \vertex (v1);
        \vertex [right=2cm of v1, dot] (v2) {};
        \vertex [above left=2cm of v1] (g_in) {\(\gamma\)};
        \vertex [below left=2cm of v1] (e_in) {\(e\)};
        \vertex [above right=2cm of v2] (a_out) {\(a\)};
        \vertex [below right=2cm of v2] (e_out) {\(e\)};
        \diagram* {
          (g_in) -- [photon] (v1),
          (e_in) -- [plain] (v1),
          (v1) -- [plain, edge label'=\(e\)] (v2),
          (v2) -- [scalar] (a_out),
          (v2) -- [plain] (e_out),
        };
        \end{feynman}
        \end{tikzpicture}
        \caption{Compton-like production}
        \label{fig:comptonlike}
    \end{subfigure}
    \hfill
    \begin{subfigure}[b]{0.45\textwidth}
        \centering
        \begin{tikzpicture}[scale=0.55, every node/.style={scale=0.85}]
        \begin{feynman}
        \vertex (v1);
        \vertex [right=2cm of v1, dot] (v2) {};
        \vertex [above left=2cm of v1] (a_in) {\(a\)};
        \vertex [below left=2cm of v1] (e_in) {\(e\)};
        \vertex [above right=2.3cm of v2] (g_out) {\(\gamma\)};
        \vertex [below right=2cm of v2] (e_out) {\(e\)};
        \diagram* {
          (a_in) -- [scalar] (v1),
          (e_in) -- [plain] (v1),
          (v1) -- [plain, edge label'=\(e\)] (v2),
          (v2) -- [photon] (g_out),
          (v2) -- [plain] (e_out),
        };
        \end{feynman}
        \end{tikzpicture}
        \caption{Inverse Compton-like scattering}
        \label{fig:inverse_compton}
    \end{subfigure}

    \vspace{1em}

    \begin{subfigure}[b]{0.45\textwidth}
        \centering
        \begin{tikzpicture}[scale=0.55, every node/.style={scale=0.85}]
        \begin{feynman}
        \vertex (a) at (-3.5, 2) {\(a\)};
        \vertex (e_in) at (-3.5, 0) {\(e\)};
        \vertex[dot] (v1) at (0, 1) {};
        \vertex (v2) at (0, -1);
        \vertex (e_out) at (3.5, -1) {\(e\)};
        \vertex (v3) at (0, -3.5);
        \vertex (N_in) at (-3.5, -3.5) {\(N,e\)};
        \vertex (N_out) at (3.5, -3.5) {\(N,e\)};
        \diagram* {
          (a) -- [scalar] (v1),
          (e_in) -- [plain] (v1),
          (v1) -- [plain, edge label=\(e\)] (v2) -- [plain] (e_out),
          (v2) -- [photon, edge label=\(\gamma\)] (v3),
          (N_in) -- [plain, thick] (v3) -- [plain, thick] (N_out)
        };
        \end{feynman}
        \end{tikzpicture}
        \caption{Axio-electric absorption}
        \label{fig:axioelectric}
    \end{subfigure}
    \hfill
    \begin{subfigure}[b]{0.45\textwidth}
        \centering
        \begin{tikzpicture}[scale=0.55, every node/.style={scale=0.85}]
        \begin{feynman}
        \vertex (a_in) {\(a\)};
        \vertex [right=2.5cm of a_in, dot] (v1) {};
        \vertex [above right=2cm of v1] (e_minus) {\(e^-\)};
        \vertex [below right=2cm of v1] (e_plus) {\(e^+\)};
        \diagram* {
          (a_in) -- [scalar] (v1),
          (v1) -- [plain] (e_minus),
          (v1) -- [plain] (e_plus),
        };
        \end{feynman}
        \end{tikzpicture}
        \caption{ALP decay into an $e^+e^-$ pair}
        \label{fig:alp_decay_ee}
    \end{subfigure}

    \caption{Production and detection processes retained for a dominant
    electron--ALP coupling.  ALPs are produced by
    $\gamma e^-\to a e^-$ and are detected through
    $a e^-\to\gamma e^-$, axio-electric absorption, or
    $a\to e^+e^-$ when $m_a>2m_e$.}
    \label{fig:compton_processes}
\end{figure}
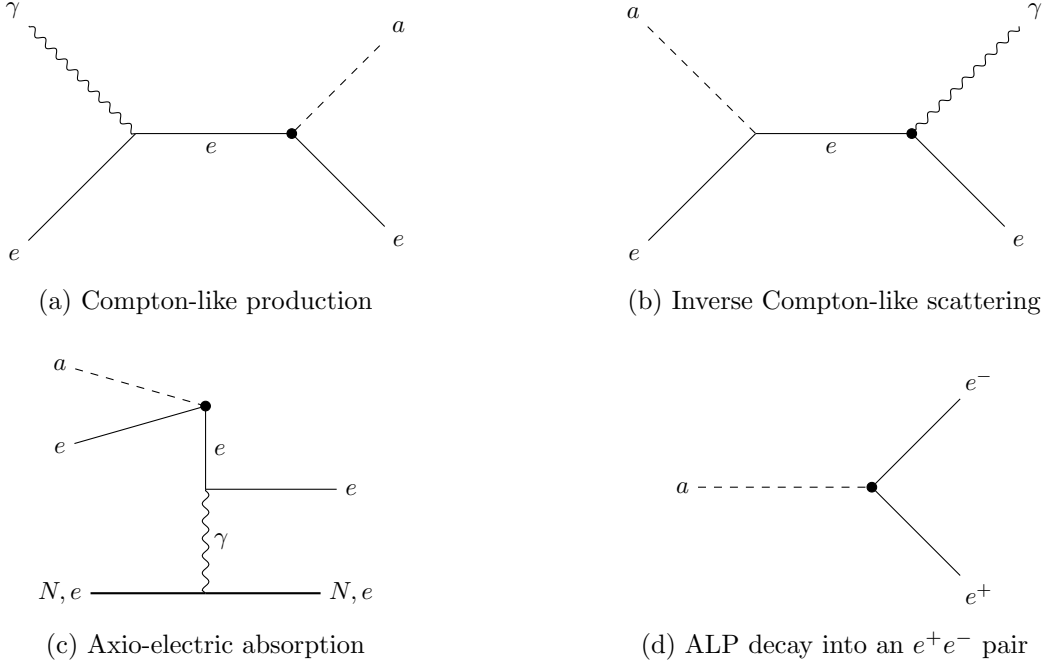

\paragraph{Reactor production and propagation}
\label{subsec:gaee_production_propagation}

Our analysis uses the common reactor photon source of
Eq.~\eqref{eq:reactor_gamma_spectrum}.  As in the photon-coupling analysis,
the production medium is represented by an effective uranium target.  In the
Compton-like process the relevant scatterers are the atomic electrons, so the
production cross section carries the uranium atomic number $Z_U=92$.  The
electrons are treated as initially free and at rest; atomic-binding effects in
the production medium are neglected.

For an incident reactor photon of energy $E'_\gamma$, the differential
cross section for $\gamma e^-\to a e^-$ is written as
~\cite{Brodsky:1986js,AristizabalSierra:2020rom}
\begin{equation}
    \frac{d\sigma_{P}^{C}}{dE_a}
    =
    \frac{Z_U\alpha g_{aee}^{2}x}
    {4(s-m_e^2)(1-x)E'_\gamma}
    \left[
    x-\frac{2m_a^2s}{(s-m_e^2)^2}
    +\frac{2m_a^2}{(s-m_e^2)^2}
    \left(
    \frac{m_e^2}{1-x}+\frac{m_a^2}{x}
    \right)
    \right],
    \label{eq:compton_production_gaee}
\end{equation}
where
\begin{equation}
    s=m_e^2+2m_eE'_\gamma,
    \qquad
    x=1-\frac{E_a}{E'_\gamma}
    +\frac{m_a^2}{2m_eE'_\gamma}.
    \label{eq:compton_variables_gaee}
\end{equation}
At fixed $E'_\gamma$ the two-body phase space restricts $x$
to~\cite{Chanda:1988}
\begin{equation}
    x_{\min,\max}
    =
    \frac{
    s-m_e^2+m_a^2
    \mp
    \sqrt{(s-m_e^2+m_a^2)^2-4sm_a^2}
    }{2s},
    \label{eq:x_limits_gaee}
\end{equation}
and the production threshold is
\begin{equation}
    E'_{\gamma,\mathrm{thr}}
    =
    m_a+\frac{m_a^2}{2m_e}.
    \label{eq:compton_threshold_gaee}
\end{equation}
The latter is particularly relevant in the MeV-mass region because the
reactor spectrum decreases exponentially with $E'_\gamma$ and terminates at
$10~\mathrm{MeV}$ in the present calculation.

The photon interaction cross section in uranium is the XCOM one of Eq.~\eqref{eq:xcom_conversion_gagg}.  Since the Compton-like contribution is negligible with respect to it over the couplings considered, we use $\sigma_{\rm Tot}^{U}\simeq\sigma_{\rm SM}^{U}$ in the production probability.

The survival-weighted differential ALP emission rate is then
\begin{equation}
    \frac{d\dot N_a^{C}}{dE_a}
    =
    P_{\rm surv}(E_a)
    \int dE'_\gamma\,
    \frac{1}{\sigma_{\rm SM}^{U}(E'_\gamma)}
    \frac{d\sigma_{P}^{C}}{dE_a}
    \frac{d\dot N_\gamma}{dE'_\gamma},
    \label{eq:alp_source_rate_gaee}
\end{equation}
where the integration is performed only over the exact kinematically allowed
incident-photon intervals implied by
Eq.~\eqref{eq:x_limits_gaee}.  As before, the
geometrical dilution $1/(4\pi L^2)$ is introduced only when computing the
detector event yield.

The decay $a\to e^+e^-$ becomes available at $m_a>2m_e$, with tree-level
width~\cite{Irastorza:2018dyq,AristizabalSierra:2020rom}
\begin{equation}
    \Gamma_{a\to e^+e^-}
    =
    \frac{g_{aee}^{2}m_a}{8\pi}
    \sqrt{1-\frac{4m_e^2}{m_a^2}}\,
    \Theta(m_a-2m_e),
    \label{eq:decay_width_gaee}
\end{equation}
where $\Theta$ denotes the Heaviside step function.  Loop-induced diphoton decays are neglected in the single-$g_{aee}$ scenario.
For $p_a=\sqrt{E_a^2-m_a^2}$, propagation from the reactor to the detector is
therefore described by
\begin{equation}
    P_{\rm surv}(E_a)
    =
    \exp\!\left[
    -\frac{Lm_a\Gamma_{a\to e^+e^-}}{p_a\hbar c}
    \right],
    \label{eq:survival_gaee}
\end{equation}
while the conditional probability for an ALP reaching the detector to decay
inside an effective path length $\ell_{\rm eff}$ is
\begin{equation}
    P_{\rm dec}^{ee}(E_a)
    =
    1-
    \exp\!\left[
    -\frac{\ell_{\rm eff}m_a\Gamma_{a\to e^+e^-}}
    {p_a\hbar c}
    \right].
    \label{eq:decay_probability_gaee}
\end{equation}
Below threshold, $P_{\rm surv}=1$ and $P_{\rm dec}^{ee}=0$ within the present
single-coupling approximation.

\paragraph{Detection mechanisms and event yields}
\label{subsec:gaee_detection}

For inverse Compton-like detection,
$a+e^-\to\gamma+e^-$, we use the differential cross section for a massive
ALP of Ref.~\cite{Avignone:1988}, in the form given in
Ref.~\cite{AristizabalSierra:2020rom},
\begin{multline}
\frac{d\sigma_D^C}{dE_\gamma}
=
\frac{Z_{\rm Ge}g_{aee}^2\alpha E_\gamma}
{4m_e^2p_a}
\left|
\frac{2(E_a+m_e-p_a\cos\theta)^2}{p_a y}
\right|
\\
\times
\left[
1+\frac{4m_e^2E_\gamma^2}{y^2}
-\frac{4m_eE_\gamma}{y}
-\frac{4m_a^2p_a^2m_eE_\gamma
(1-\cos^2\theta)}{y^3}
\right],
\label{eq:inverse_compton_gaee}
\end{multline}
where $Z_{\rm Ge}=32$, $\theta$ is the angle between the incident ALP
and the outgoing photon, and
\begin{equation}
    y=2m_eE_a+m_a^2,
    \qquad
    \cos\theta=
    \frac{E_a+m_e-y/(2E_\gamma)}{p_a}.
\end{equation}
For each $E_a$ we integrate over the exact final-state photon interval
\begin{align}
E_{\gamma}^{\min}
&=
\max\!\left[
25~\mathrm{keV},
\frac{y}{2(E_a+m_e+p_a)}
\right],
\\
E_{\gamma}^{\max}
&=
\min\!\left[
10~\mathrm{MeV},
\frac{y}{2(E_a+m_e-p_a)}
\right],
\label{eq:inverse_compton_bounds_gaee}
\end{align}
and define
\begin{equation}
    \sigma_D^C(E_a)
    =
    \int_{E_\gamma^{\min}}^{E_\gamma^{\max}}
    \frac{d\sigma_D^C}{dE_\gamma}\,dE_\gamma.
    \label{eq:inverse_compton_total_gaee}
\end{equation}

The second scattering channel is axio-electric absorption.  Its cross section
is related to the ordinary photoelectric cross section of germanium
by~\cite{Dimopoulos:1986,Avignone:1987,Pospelov:2008,Derevianko:2010}
\begin{equation}
    \sigma_D^A(E_a)
    =
    \frac{g_{aee}^{2}}{\beta}
    \frac{3E_a^2}{16\pi\alpha m_e^2}
    \left(1-\frac{\beta^{2/3}}{3}\right)
    \sigma_{\rm PE}^{\rm Ge}(E_a),
    \qquad
    \beta=\frac{p_a}{E_a},
    \label{eq:axioelectric_gaee}
\end{equation}
where $\sigma_{\rm PE}^{\rm Ge}$ is taken from the germanium XCOM table.

The corresponding scattering event yields are
\begin{align}
    N_{\rm IC}
    &=
    \frac{M_{\rm det}N_T\Delta t}{4\pi L^2}
    \int dE_a\,
    \frac{d\dot N_a^C}{dE_a}\,
    \sigma_D^C(E_a),
    \label{eq:inverse_compton_yield_gaee}
    \\
    N_{\rm AE}
    &=
    \frac{M_{\rm det}N_T\Delta t}{4\pi L^2}
    \int dE_a\,
    \frac{d\dot N_a^C}{dE_a}\,
    \sigma_D^A(E_a),
    \label{eq:axioelectric_yield_gaee}
\end{align}
where $N_T$ is the number of germanium atoms per unit detector mass, as
defined in Sec.~\ref{subsec:gagg_detection}.

For $m_a>2m_e$, ALPs can also be detected through an in-detector
electron--positron decay.  We use the same effective decay geometry,
characterized by $A_{\rm proj}$ and $\ell_{\rm eff}$, introduced in
Sec.~\ref{subsec:gagg_detection}.  The corresponding
yield is
\begin{equation}
    N_{ee}
    =
    \frac{A_{\rm proj}\Delta t}{4\pi L^2}
    \int dE_a\,
    \frac{d\dot N_a^C}{dE_a}\,
    P_{\rm dec}^{ee}(E_a).
    \label{eq:ee_decay_yield_gaee}
\end{equation}

The total single-coupling electron--ALP signal is therefore
\begin{equation}
    S_{aee}
    =
    N_{\rm IC}+N_{\rm AE}+N_{ee}.
    \label{eq:total_signal_gaee}
\end{equation}

\subsubsection{Nucleon--ALP coupling}
\label{subsec:gann}

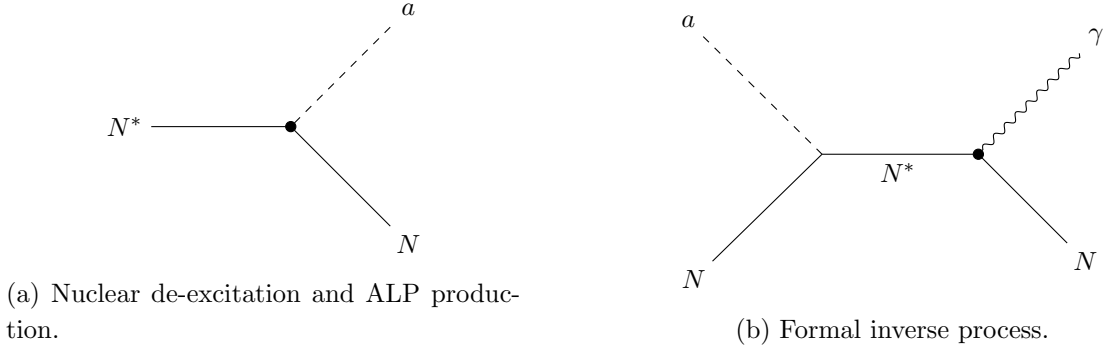
\begin{figure}[tbp]
    \centering
    \begin{subfigure}[b]{0.45\textwidth}
        \centering
        \begin{tikzpicture}[scale=0.75, every node/.style={scale=0.90}]
        \begin{feynman}
            \vertex (Nstar) {\(N^{*}\)};
            \vertex [right=2.2cm of Nstar, dot] (v1) {};
            \vertex [above right=2.2cm of v1] (aout) {\(a\)};
            \vertex [below right=2.2cm of v1] (Nout) {\(N\)};
            \diagram* {
                (Nstar) -- [plain] (v1),
                (v1) -- [scalar] (aout),
                (v1) -- [plain] (Nout),
            };
        \end{feynman}
        \end{tikzpicture}
        \caption{Nuclear de-excitation and ALP production.}
        \label{fig:gann_production}
    \end{subfigure}
    \hfill
    \begin{subfigure}[b]{0.45\textwidth}
        \centering
        \begin{tikzpicture}[scale=0.65, every node/.style={scale=0.90}]
        \begin{feynman}
            \vertex (v1);
            \vertex [right=2.0cm of v1, dot] (v2) {};
            \vertex [above left=2.2cm of v1] (ain) {\(a\)};
            \vertex [below left=2.0cm of v1] (Nin) {\(N\)};
            \vertex [above right=2.2cm of v2] (gout) {\(\gamma\)};
            \vertex [below right=2.0cm of v2] (Nout) {\(N\)};
            \diagram* {
                (ain) -- [scalar] (v1),
                (Nin) -- [plain] (v1),
                (v1) -- [plain, edge label'=\(N^{*}\)] (v2),
                (v2) -- [photon] (gout),
                (v2) -- [plain] (Nout),
            };
        \end{feynman}
        \end{tikzpicture}
        \caption{Formal inverse process.}
        \label{fig:gann_inverse}
    \end{subfigure}
    \caption{Nuclear processes associated with the nucleon--ALP coupling.
    Panel (a) shows the production mechanism used in this work,
    $N^{*}\to N+a$.  Panel (b) shows the formal inverse reaction
    $a+N\to N^{*}\to N+\gamma$, which is not used for detection because the
    germanium target does not provide the resonant isotopes and transitions
    required to absorb the reactor ALP lines.}
    \label{fig:gann_processes}
\end{figure}

We finally consider ALP production through the nucleon interactions already introduced in Eq.~\eqref{eq:interactions-ALP}.  In contrast to Primakoff and Compton-like production, this channel is generated by discrete magnetic nuclear transitions rather than by the continuous reactor photon spectrum of Eq.~\eqref{eq:reactor_gamma_spectrum}.  Following neutron capture or the production of excited fission fragments, an excited nuclear state $N^*$ may emit an ALP instead of a photon.  Pseudoscalar emission competes with magnetic
$MJ$ transitions, whereas electric transitions do not contribute through the nucleon coupling~\cite{Weinberg:1978,DonnellyAxions:1978,AristizabalSierra:2020rom}.  Figure~\ref{fig:gann_processes} shows the production process $N^*\to N+a$ and its formal inverse; the latter is displayed only for completeness and is not used as a detection channel here.  The magnetic transitions and their photon yields are described in
Sec.~\ref{subsec:reactor_lines}.

\paragraph{ALP-to-photon branching ratios}
\label{par:gann_branching}

For a transition $i$, ALP production is determined by the ratio of the ALP and
photon partial widths, $\Gamma_i^{a}/\Gamma_i^{\gamma}$.  We distinguish these
nuclear-transition widths from the total particle decay width
$\Gamma_a^{\mathrm{tot}}$, which controls propagation.  For the isovector
$M1$ transition $p(n,\gamma)d$, the ratio is
~\cite{Barroso:1981,AristizabalSierra:2020rom}
\begin{equation}
    \left(\frac{\Gamma_i^{a}}{\Gamma_i^{\gamma}}\right)_{pn}
    =\frac{1}{2\pi\alpha}
    \left(\frac{p_a}{p_\gamma}\right)^3
    \left(\frac{g_{ann}^{(1)}}{\mu_1}\right)^2,
    \label{eq:gann_branching_pn}
\end{equation}
where $\mu_1=\mu_p-\mu_n=4.71$ is the isovector nuclear magnetic
moment in units of the nuclear magneton~\cite{Avignone:1987}.  Neglecting the very small nuclear
recoil, $p_\gamma\simeq E_i$ and
$p_a=\sqrt{E_i^2-m_a^2}$.  Therefore, emission from line $i$ is allowed only
for $m_a<E_i$.

For a general magnetic transition of multipolarity $MJ$, the long-wavelength
result is~\cite{Avignone:1987,TEXONO:2006axion,AristizabalSierra:2020rom}
\begin{equation}
    \left(\frac{\Gamma_i^{a}}{\Gamma_i^{\gamma}}\right)_{MJ}
    =\frac{1}{\pi\alpha}\frac{1}{1+\delta_i^2}
    \frac{J}{J+1}
    \left(\frac{p_a}{p_\gamma}\right)^{2J+1}
    \left[
    \frac{g_{ann}^{(0)}\kappa_i+g_{ann}^{(1)}}
    {(\mu_0-\tfrac{1}{2})\kappa_i+(\mu_1+\eta_i)}
    \right]^2,
    \label{eq:gann_branching_MJ}
\end{equation}
where $J$ is the magnetic multipole order and
$\mu_0=\mu_p+\mu_n=0.88$ is the isoscalar nuclear magnetic moment, also in
nuclear-magneton units.  The parameter $\delta_i$ is the electric-to-
magnetic multipole mixing ratio, while $\eta_i$ and $\kappa_i$ encode ratios
of nuclear spin and orbital matrix elements~\cite{Avignone:1987}.  For the
predominantly $M1$ $^{7}\mathrm{Li}^{*}$ transition, we use $\delta_i\ll1$,
$\eta_i=0.5$, and $\kappa_i=1.0$, as calculated in Ref.~\cite{Krcmar:2001}
and adopted by TEXONO~\cite{TEXONO:2006axion}.  For the $M4$ lines no
calculations are available; TEXONO estimated them from the results for
$^{65}\mathrm{Cu}$, $^{57}\mathrm{Fe}$, $^{55}\mathrm{Mn}$ and
$^{23}\mathrm{Na}$ of Refs.~\cite{Avignone:1987,HaxtonLee:1991}, obtaining
$\delta_i=0$, $\eta_i=1$, and $\kappa_i=-3$ for $^{91}\mathrm{Y}^{*}$ and
$^{97}\mathrm{Nb}^{*}$, and $\delta_i=0$, $\eta_i=-1$, and $\kappa_i=1$ for
$^{135}\mathrm{Xe}^{*}$ and $^{137}\mathrm{Ba}^{*}$~\cite{TEXONO:2006axion}.
We adopt the same values as Ref.~\cite{AristizabalSierra:2020rom}.
These nuclear-structure inputs are adopted as fixed benchmark values, and
their uncertainties are not propagated as nuisance parameters.

\paragraph{Monochromatic ALP source and propagation}
\label{par:gann_source}

The transition energy $E_i$ is fixed by the nuclear level spacing.  After
neglecting nuclear recoil and the intrinsic line width, the corresponding ALP
source is treated as monochromatic.  The photon yields $\Phi_{\gamma,i}$
and the fission rate $R_f(P)$ are those of Sec.~\ref{subsec:reactor_lines},
Eq.~\eqref{eq:gann_fission_rate}.

Defining
\begin{equation}
    \mathcal{B}_i^a
    \equiv \frac{\Gamma_i^{a}}{\Gamma_i^{\gamma}},
    \label{eq:gann_Ba_definition}
\end{equation}
the survival-weighted differential ALP emission rate from transition $i$ is
\begin{equation}
    \frac{d\dot N_{a,i}^{\mathrm{MT}}}{dE_a}
    =R_f(P)\,\Phi_{\gamma,i}\,\mathcal{B}_i^a\,
    P_{\mathrm{surv}}(E_i)\,
    \delta(E_a-E_i)\,\Theta(E_i-m_a).
    \label{eq:gann_monochromatic_source}
\end{equation}
Here the superscript ``MT'' denotes production through a magnetic nuclear transition.  Equation~\eqref{eq:gann_monochromatic_source} uses the weak-branching approximation $\Gamma_i^a\ll\Gamma_i^\gamma$, which is valid throughout the parameter region considered.  In the mixed-coupling projections below, propagation is evaluated with the decay width associated with the detection coupling: $\Gamma_{a\to\gamma\gamma}$ for photon-mediated
detection and $\Gamma_{a\to e^+e^-}$ for electron-mediated detection.  The
corresponding survival probabilities are those defined in
Eqs.~\eqref{eq:survival_gagg} and \eqref{eq:survival_gaee}, respectively.  In
the long-lived region both approach unity.

\paragraph{Detection in germanium}
\label{par:gann_detection}

Detection through resonant nuclear excitation would require a target
containing the appropriate isotopes and matching transitions.  Since the
germanium detectors do not satisfy this requirement for the six
reactor lines considered here, the nuclear-production channel cannot constrain
$g_{ann}^{(0,1)}$ by itself.  Instead, the produced ALPs are detected through a
second coupling: inverse Primakoff conversion for $g_{a\gamma\gamma}$, or
inverse Compton-like scattering and axio-electric absorption for $g_{aee}$.
In-detector decays into photons or electron--positron pairs can also contribute
when they are kinematically allowed.  The resulting analysis therefore probes
coupling pairs rather than an isolated nucleon coupling.

For compactness, define the survival-weighted line rate
\begin{equation}
    \varphi_{a,i}
    =R_f(P)\,\Phi_{\gamma,i}\,\mathcal{B}_i^a\,
    P_{\mathrm{surv}}(E_i)\,\Theta(E_i-m_a),
    \label{eq:gann_line_rate}
\end{equation}
which has units of $\mathrm{s}^{-1}$.  Using the detection cross sections
defined in the preceding subsections, the expected number of events from line
$i$ is
\begin{align}
    N_{P,i}^{\mathrm{MT}}
    &=\frac{M_{\mathrm{det}}N_T\Delta t}{4\pi L^2}
      \varphi_{a,i}\,\sigma_D^P(E_i),
      \label{eq:gann_yield_primakoff}\\
    N_{C,i}^{\mathrm{MT}}
    &=\frac{M_{\mathrm{det}}N_T\Delta t}{4\pi L^2}
      \varphi_{a,i}\,\sigma_D^C(E_i),
      \label{eq:gann_yield_compton}\\
    N_{A,i}^{\mathrm{MT}}
    &=\frac{M_{\mathrm{det}}N_T\Delta t}{4\pi L^2}
      \varphi_{a,i}\,\sigma_D^A(E_i),
      \label{eq:gann_yield_axioelectric}
\end{align}
where the inverse-Compton cross section $\sigma_D^C(E_i)$ is understood to be
integrated over the accepted final-state energy interval.

For a visible decay mode $f$, the corresponding yield is
\begin{equation}
    N_{\mathrm{dec},i}^{\mathrm{MT},f}
    =\frac{A_{\rm proj}\Delta t}{4\pi L^2}
     \varphi_{a,i}\,P_{\mathrm{dec}}^{f}(E_i),
    \label{eq:gann_yield_decay}
\end{equation}
with
\begin{equation}
    P_{\mathrm{dec}}^{f}(E_i)
    =\frac{\Gamma_{a\to f}}{\Gamma_a^{\mathrm{tot}}}
    \left[1-
    \exp\!\left(-\frac{\ell_{\rm eff}m_a\Gamma_a^{\mathrm{tot}}}
    {p_a\hbar c}\right)\right].
    \label{eq:gann_decay_probability}
\end{equation}
Here $p_a=\sqrt{E_i^2-m_a^2}$ and $\Gamma_a^{\rm tot}$ denotes the
total ALP decay width relevant for the mixed-coupling scenario.  The decay
term uses the same $A_{\rm proj}$ and $\ell_{\rm eff}$ geometries adopted
in the single-coupling photon and electron analyses.  If
only one decay mode is open, the branching-fraction prefactor in
Eq.~\eqref{eq:gann_decay_probability} equals unity.

The six transitions above are retained to characterize the magnetic-line
reactor source.  For the mixed-coupling sensitivity projections in
Sec.~\ref{subsec:alp_discussion}, however, we follow
Ref.~\cite{AristizabalSierra:2020rom} and retain only the dominant isovector
$p(n,\gamma)d$ contribution.  We furthermore adopt
\begin{equation}
    g_{ann}^{(0)}=0,
    \label{eq:gann_isovector_benchmark}
\end{equation}
so that nuclear production is controlled by $g_{ann}^{(1)}$.  In the
weak-coupling, long-lived limit, inverse-Primakoff detection then scales as
$(g_{ann}^{(1)}g_{a\gamma\gamma})^2$, while inverse-Compton-like and
axio-electric detection scale as $(g_{ann}^{(1)}g_{aee})^2$.  Away from this
limit, the propagation and decay probabilities retain a separate dependence
on the detection coupling.

\section{Application to CONUS+ and CONUS100}
\label{sec:conus}

The framework of Secs.~\ref{sec:reactor_source}--\ref{Model} is now
applied to the CONUS+ experiment and to the proposed CONUS100 upgrade.

\subsection{Experimental setup}
\label{subsec:conus_setup}

\begin{figure}[H]
    \centering
    \includegraphics[width=0.5\linewidth]{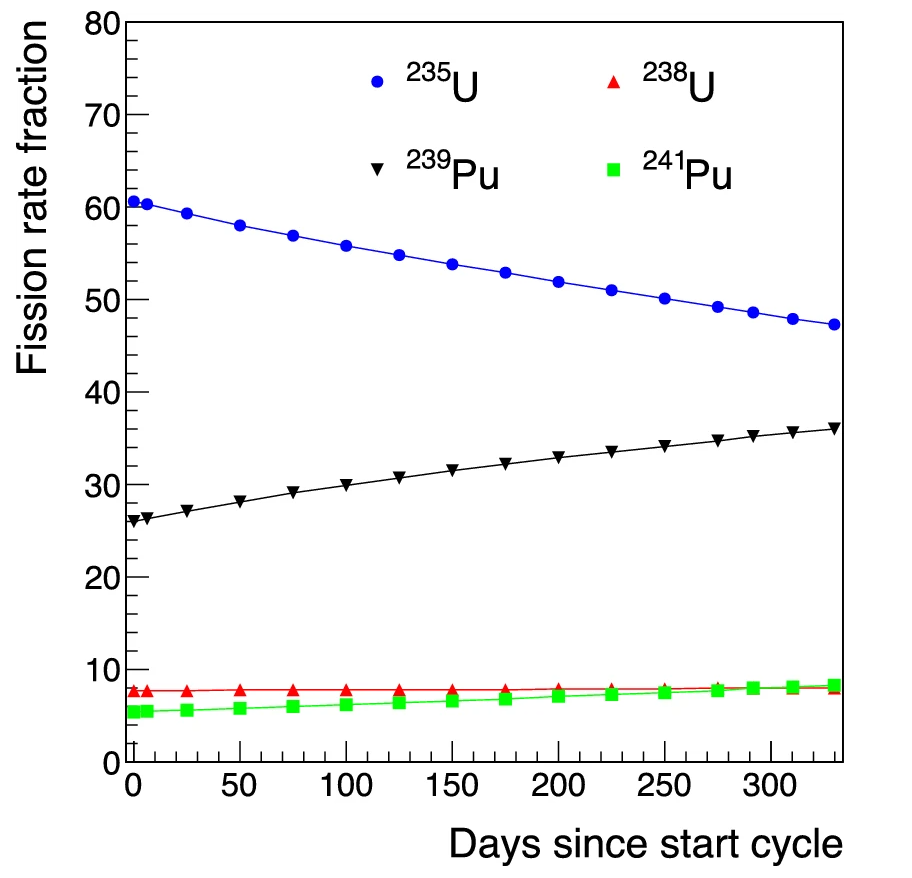}
    \caption{Exemplary average fission rate fractions of the four dominant fissile actinides as function of one typical reactor cycle evolution at KKL. Retrieved from \cite{ConusPlus:2024}.}
    \label{fig:fission}
\end{figure}

CONUS+ was installed in 2023 at the Leibstadt nuclear power plant (KKL), Switzerland, as the successor of the CONUS experiment at the Brokdorf reactor (KBR), Germany.  Its primary goal is the study of coherent elastic neutrino--nucleus scattering (CE$\nu$NS) and the search for deviations from the SM prediction, with the boiling-water reactor as antineutrino source.  The same environment is favourable for dark photon and ALP searches: the reactor operates at a maximum thermal power of $3.6~\mathrm{GW}$, and the antineutrino flux at the detector site, $1.45\times10^{13}~\mathrm{cm^{-2}\,s^{-1}}$, reflects the high fission rate (Fig.~\ref{fig:fission}) and hence the intense photon emission of the core.

In the CONUS+ area, the average radiation dose is below $1~\mu\mathrm{Sv\,h^{-1}}$ and the average overburden is $7.4$~m of water equivalent \cite{ConusPlus:2024}.

The detector setup is situated at a distance of $\sim20.7$~m from the reactor core. It consists of 4 High Purity Germanium (HPGe) detectors C2, C3, C4, C5, each having 1 kg mass and resulting in an active mass of $0.92$--$0.95$~kg per detector.  The decreasing trigger efficiency towards very low energies is substantially compensated by the small point contact size design choice of the P-type Point Contact (PPC) detectors. This not only decreases the capacitance, but also features a highly steep field gradient near the point contact, improving further the pulse shape discrimination. Furthermore, in order to overcome the undesired effects of thermal excitation to a good extent, the detectors are kept inside Copper cryostats connected to cryocoolers outside the shield. The high-purity crystal choice is also crucial for the minimization of the leakage current~\cite{Ackermann2025}.

\begin{figure}[H]
    \centering
    \includegraphics[width=0.5\linewidth]{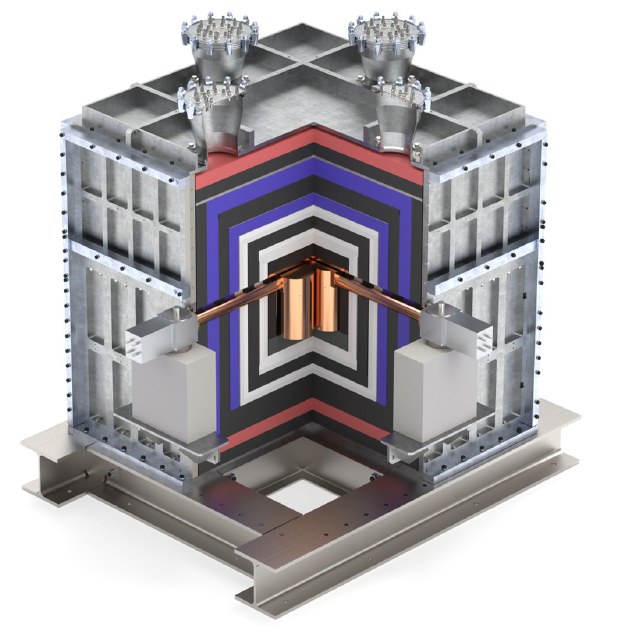}
    \caption{CONUS+ setup: Adapted from \cite{ConusPlus:2024}.}
    \label{fig:shield}
\end{figure}

Outside the detectors, an onion-like structure of shield layers as in Figure \ref{fig:shield} is present for mitigation of several background components:
\begin{enumerate}[label=(\roman*)]
    \item Four layers of Pb as passive shield, 
    \item Two plates of polyethylene and two layers of B-doped polyethylene as neutron shield, and 
    \item Two layers of plastic scintillator plates as active $\mu$ veto
\end{enumerate}

This whole setup is housed inside a steel frame accounting for the strict earthquake requirements of the site. Radon diffusion is reduced by the radon-seal tape at the frame edges.

\textbf{Detector Performance}

\begin{figure}[H]
    \centering
    \includegraphics[width=0.5\linewidth]{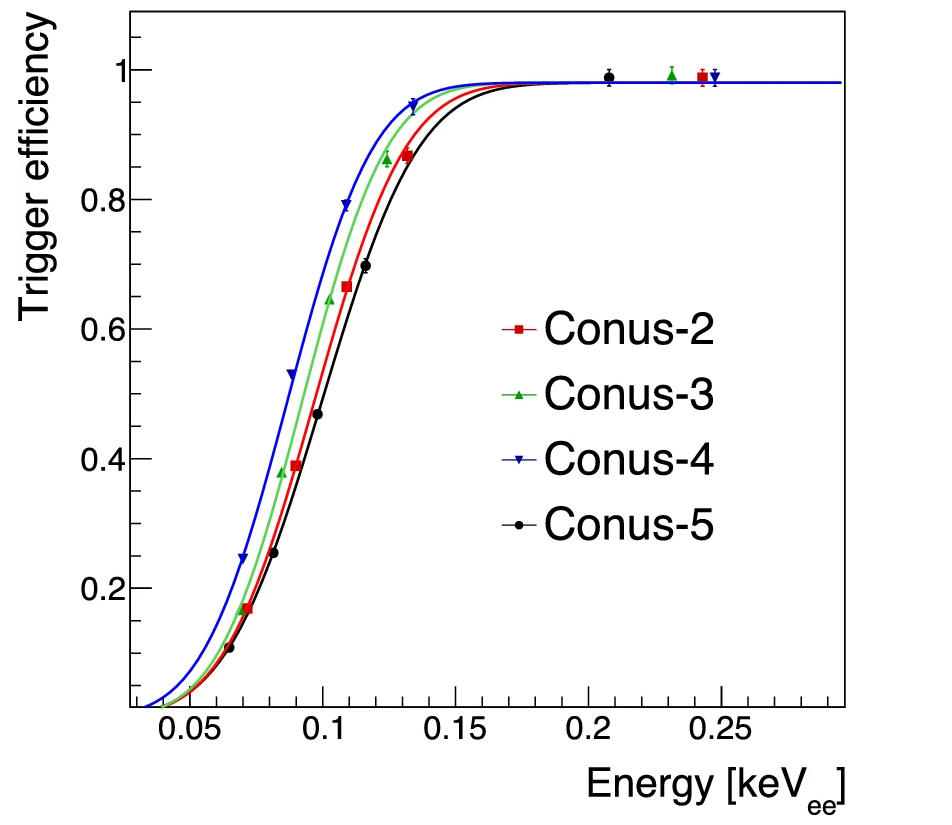}
    \caption{Trigger efficiency vs. energy curves for each CONUS+ detector. Adapted from \cite{ConusPlus:2024}.}
    \label{fig:trigeff}
\end{figure}

The resolution and trigger efficiency performances of each CONUS+ detector are given respectively in Table \ref{tab:reso} and Figure \ref{fig:trigeff}.

\begin{table}[H]
    \centering
    \begin{tabular}{lcc}
        \hline
        Detector & Pulser resolution \\
        \hline
        C2 & $48 \pm 1$ $eV_{ee}$  \\
        C3 & $47 \pm 1$ $eV_{ee}$ \\
        C4 & $47 \pm 1$ $eV_{ee}$  \\
        C5 & $47 \pm 1$ $eV_{ee}$ \\
        \hline
    \end{tabular}
    \caption{Pulser resolution measurement results of each CONUS+ detector at the KKL site. Adapted from \cite{ConusPlus:2024}.}
    \label{tab:reso}
\end{table}

\begin{figure}[H]
    \centering
    \includegraphics[width=0.5\linewidth]{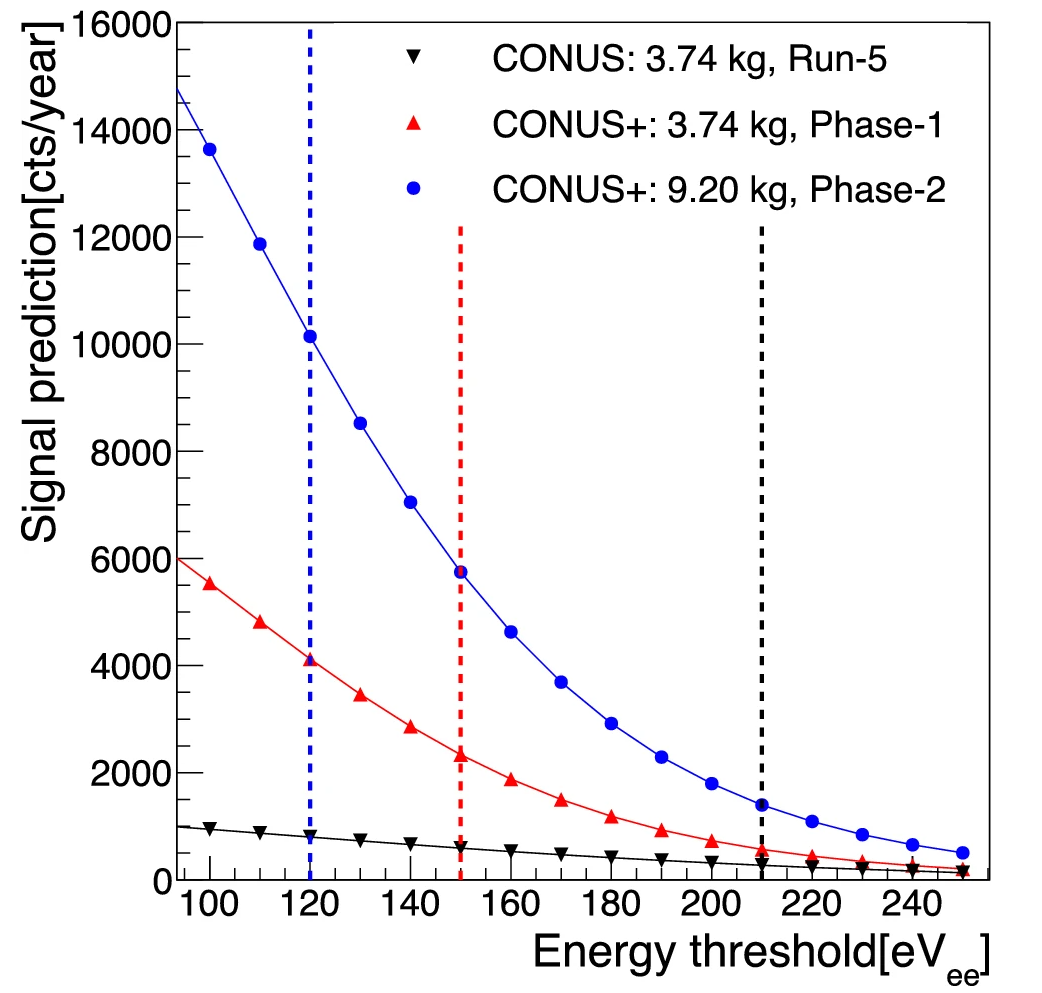}
    \caption{$CE\nu NS$ signal prediction vs. energy threshold for CONUS Run-5, CONUS+ Phase 1, and CONUS+ Phase 2. Adapted from \cite{ConusPlus:2024}.}
    \label{fig:massnthreshold}
\end{figure}

The expected CE$\nu$NS signal as a function of the energy threshold for the different CONUS and CONUS+ phases is shown in Fig.~\ref{fig:massnthreshold}.

CONUS100 is a proposed upgrade of the CONUS+ technology to a $100$~kg germanium array~\cite{Conus100:2025}.  As benchmark we take $P=3.5$~GW, $L=20$~m and $M_{\rm det}=100$~kg.  For the dark photon and ALP projections its readout is assumed to extend to $10$~MeV in deposited energy, while CONUS+ keeps its present $0.6$~MeV DAQ range.  The projected CE$\nu$NS threshold of $120$~eV is not relevant here, since the signals considered lie in the keV--MeV range.

For the ALP projections based on Sec.~\ref{Model}, we use the common
experimental benchmarks summarized in Table~\ref{tab:alp_benchmarks}.  The
quoted exposures are full-power-equivalent (FPE) exposures; reactor outages,
dead time, and detector efficiencies are treated separately as analysis
assumptions rather than as experimental parameters.  In particular, the
live-time fraction of CONUS+, $0.88$ (muon-veto dead time), is not applied in the dark photon and ALP projections.  For CONUS+ we keep the
present $0.6~\mathrm{MeV}$ DAQ range, i.e.\ unit efficiency between
$25~\mathrm{keV}$ and $0.6~\mathrm{MeV}$, since the published readout
(channels up to $35~\mathrm{keV_{ee}}$ and $600~\mathrm{keV_{ee}}$) is
optimised for CE$\nu$NS~\cite{ConusPlus:2024}.  For CONUS100 the ALP
projections instead assume that the readout is extended to
$10~\mathrm{MeV}$, with unit efficiency between $25~\mathrm{keV}$ and
$10~\mathrm{MeV}$, since the ALP signals extend to MeV energies.  The same
windows are used for the dark photon search, with a lower edge of
$0.2~\mathrm{MeV}$
(Secs.~\ref{subsec:dp_discussion} and \ref{subsec:alp_discussion}).  Including the live-time fraction would
weaken the projected couplings by only $1$--$3\%$.  With
Eq.~\eqref{eq:gann_fission_rate}, the fission rates are
$R_f=1.1232\times10^{20}~\mathrm{s}^{-1}$ for KKL ($P=3.6~\mathrm{GW}$) and
$1.092\times10^{20}~\mathrm{s}^{-1}$ for the CONUS100 benchmark
($P=3.5~\mathrm{GW}$).

\begin{table}[t]
    \centering
    \begin{tabular}{lcc}
        \hline
        Parameter & CONUS+ & CONUS100 \\
        \hline
        Reactor thermal power $P$ & $3.6~\mathrm{GW}$ & $3.5~\mathrm{GW}$ \\
        Reactor--detector distance $L$ & $20.7~\mathrm{m}$ & $20~\mathrm{m}$ \\
        Ge target mass $M_{\rm det}$ & $3.74~\mathrm{kg}$ & $100~\mathrm{kg}$ \\
        Exposure $\Delta t$ & $1~\mathrm{FPE~yr}$ & $5~\mathrm{FPE~yr}$ \\
        Accepted energy window & $25~\mathrm{keV}$ ($0.2~\mathrm{MeV}$)--$0.6~\mathrm{MeV}$ & $25~\mathrm{keV}$ ($0.2~\mathrm{MeV}$)--$10~\mathrm{MeV}$ \\
        \hline
    \end{tabular}
    \caption{Experimental benchmarks adopted for the dark photon and ALPs
    projections~\cite{ConusPlus:2024,Conus100:2025}.  The CONUS+ entries follow the present detector configuration,
    while the CONUS100 entries correspond to the proposed $100$~kg benchmark.
   The lower edge of the energy window is $25~\mathrm{keV}$ for ALPs and $0.2~\mathrm{MeV}$ (in parentheses) for dark photons.}
    \label{tab:alp_benchmarks}
\end{table}

\subsection{Effective detector geometry}
\label{subsec:conus_geometry}

The decay yields of Eqs.~\eqref{eq:diphoton_yield_gagg},
\eqref{eq:ee_decay_yield_gaee} and \eqref{eq:gann_yield_decay} require an
effective projected area $A_{\rm proj}$ and mean chord $\ell_{\rm eff}$ of the
germanium target.
For CONUS+, the four refurbished HPGe detectors are modeled as four
cylindrical crystals of height and diameter $62\,\mathrm{mm}$
~\cite{ConusPlus:2024}. Since the orientation of the crystals relative to the reactor is not modelled, we use the orientation-averaged
projected area of a convex body, $\langle A_{\rm proj}\rangle=\mathcal{S}/4$, where $\mathcal{S}$ is the crystal surface area, and sum
the four crystals without overlap. This gives
\begin{equation}
A_{\rm proj}^{\rm C+}=181.14~\mathrm{cm}^2,
\qquad
\ell_{\rm eff}^{\rm C+}=4.13~\mathrm{cm},
\end{equation}
with $A_{\rm proj}\ell_{\rm eff}$ equal to the modeled Ge volume.  The
inverse-Primakoff normalization uses the measured active mass, whereas the
decay term uses the physical crystal dimensions as an effective geometrical
interception proxy.

For CONUS100 no final crystal layout is presently specified in the proposal
~\cite{Conus100:2025}. We therefore use a compact cube having the
volume of $100\,\mathrm{kg}$ of Ge. With the standard Ge density this gives
a side length $26.58\,\mathrm{cm}$ and hence
\begin{equation}
A_{\rm proj}^{\rm C100}=706.69~\mathrm{cm}^2,
\qquad
\ell_{\rm eff}^{\rm C100}=26.58~\mathrm{cm}.
\end{equation}
This CONUS100 geometry is a projection benchmark, not an official detector
dimension. For the cube we use the face-on projection, whereas for CONUS+ we
use the orientation average; this difference has no effect on the results.
An ALP that reaches the detector has a decay length much larger than
$\ell_{\rm eff}$, since otherwise it would have decayed along the $20$~m
baseline, so that $P_{\rm dec}\simeq\ell_{\rm eff}/\lambda_{\rm dec}$ and
the decay yield depends only on the product $A_{\rm proj}\ell_{\rm eff}$,
i.e.\ on the detector volume.

The modeled CONUS+ volume, $749~\mathrm{cm^3}$, corresponds to
$3.99~\mathrm{kg}$ of Ge, i.e.\ $6.6\%$ above the active mass used for the
scattering yields; restricting the decays to the active volume would lower the
decay yields by this factor and weaken the decay-dominated reach by about
$3\%$.  Note also that the cube of Ref.~\cite{AristizabalSierra:2020rom}
($L_{\rm det}=50~\mathrm{cm}$ for $10~\mathrm{kg}$ of Ge) has a volume much
larger than that of the detector, so the decay rates of that reference are not
directly comparable with the ones derived here.

\subsection{Expected event rates}
\label{subsec:conus_rates}

\begin{figure}
    \centering
    \includegraphics[width=1\textwidth]{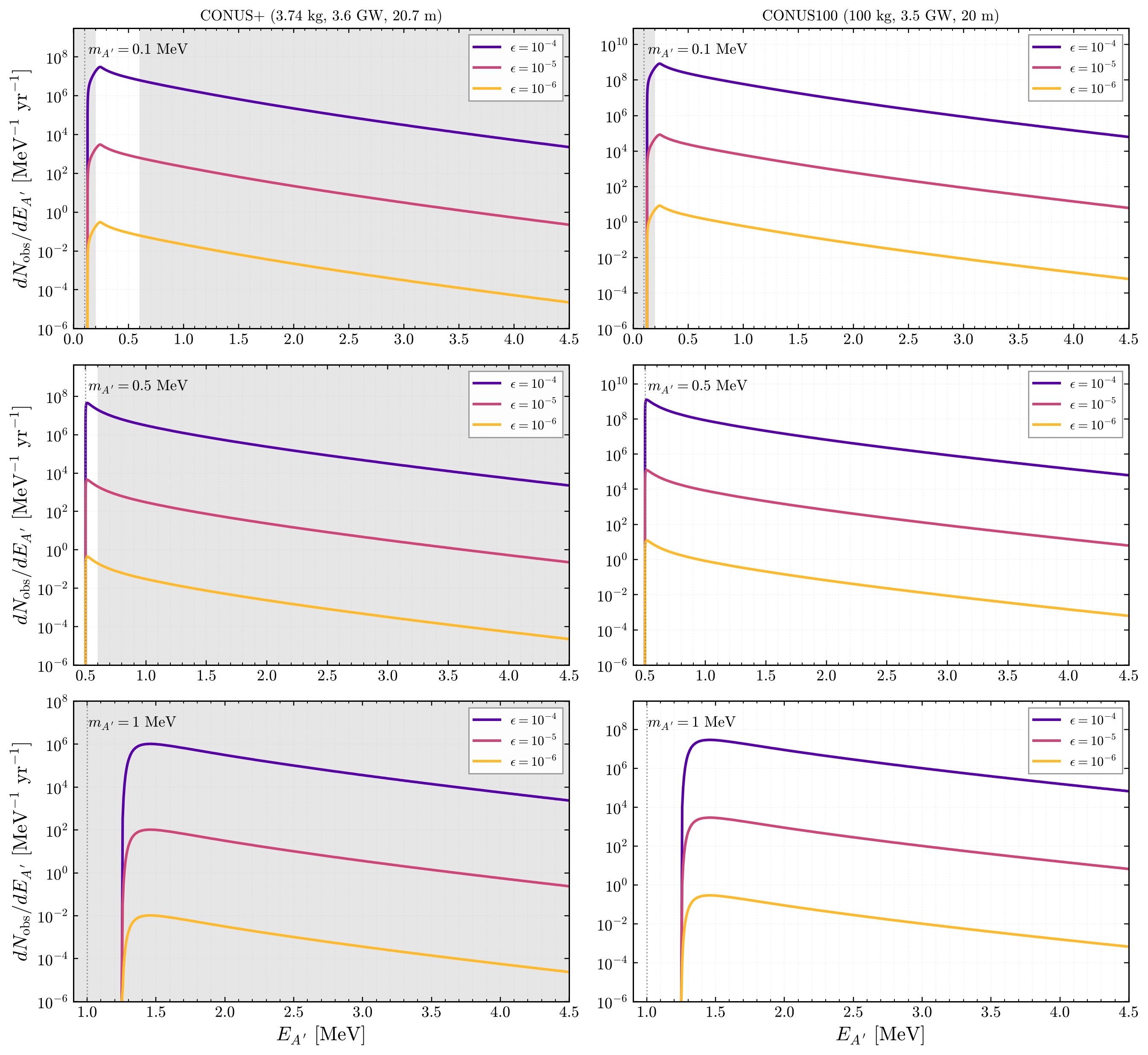}
    \caption{Differential dark photon event rates, $dN_{\rm obs}/dE_{A'}$, per MeV and per full-power year, from inverse Compton-like absorption $A'e^-\to\gamma e^-$ in CONUS+ (left) and CONUS100 (right) for $m_{A'}=0.1$, $0.5$ and $1~\mathrm{MeV}$ (rows) and $\epsilon=10^{-4}$, $10^{-5}$ and $10^{-6}$ (colours).  No efficiency or energy window is applied; the shaded areas are outside the accepted windows, below $0.2~\mathrm{MeV}$ and, for CONUS+, above $0.6~\mathrm{MeV}$.  The dotted lines mark $E_{A'}=m_{A'}$.}
\label{fig:dp_rates_conus}
    \label{fig:DP_rates_grid}
\end{figure}

\begin{figure}
    \centering
    \includegraphics[width=1\linewidth]{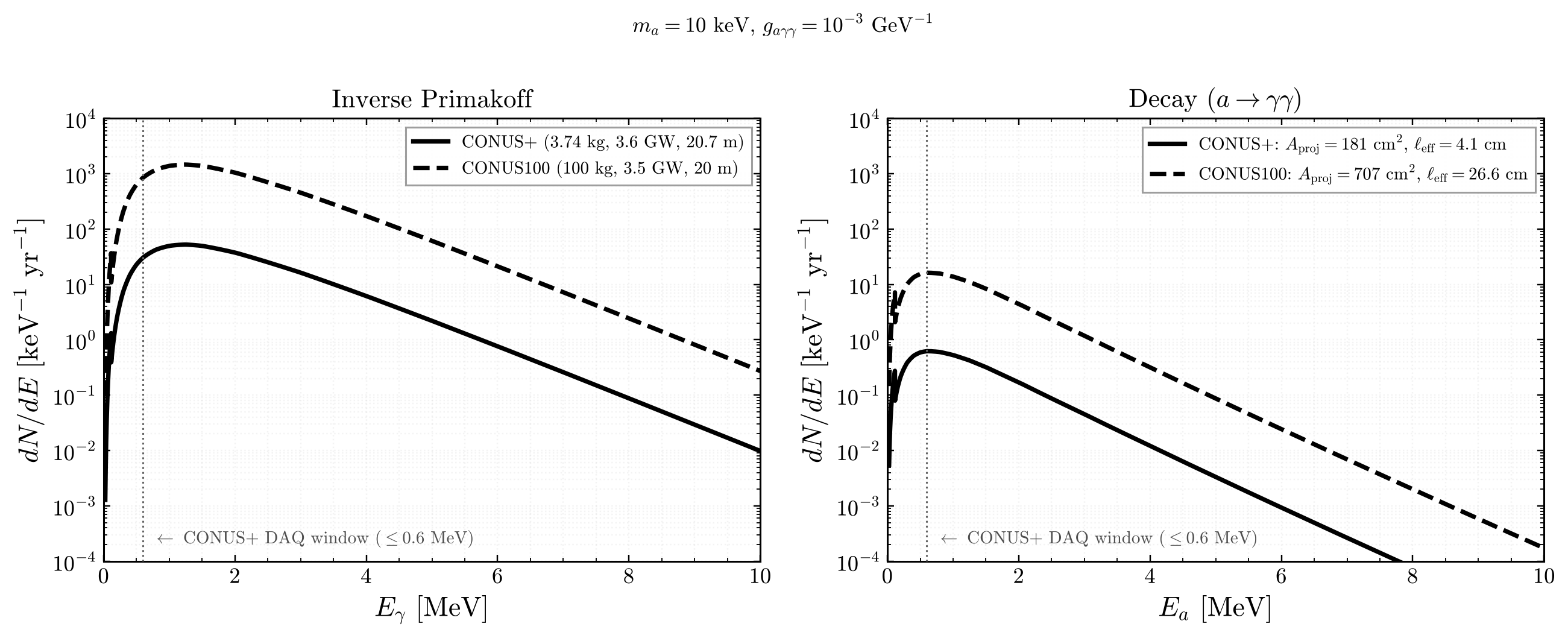}
    \caption{Differential event rates per keV and per full-power year for
    $m_a=10~\mathrm{keV}$ and $g_{a\gamma\gamma}=10^{-3}~\mathrm{GeV}^{-1}$:
    inverse Primakoff scattering (left) and $a\to\gamma\gamma$ decays inside
    the detector (right), for CONUS+ (solid) and CONUS100 (dashed).  No
    efficiency or energy window is applied; the dotted line marks the upper
    end of the CONUS+ DAQ window, $0.6~\mathrm{MeV}$.}
    \label{fig:conus_alps}
\end{figure}

\begin{figure}
    \centering
    \includegraphics[width=0.88\linewidth]{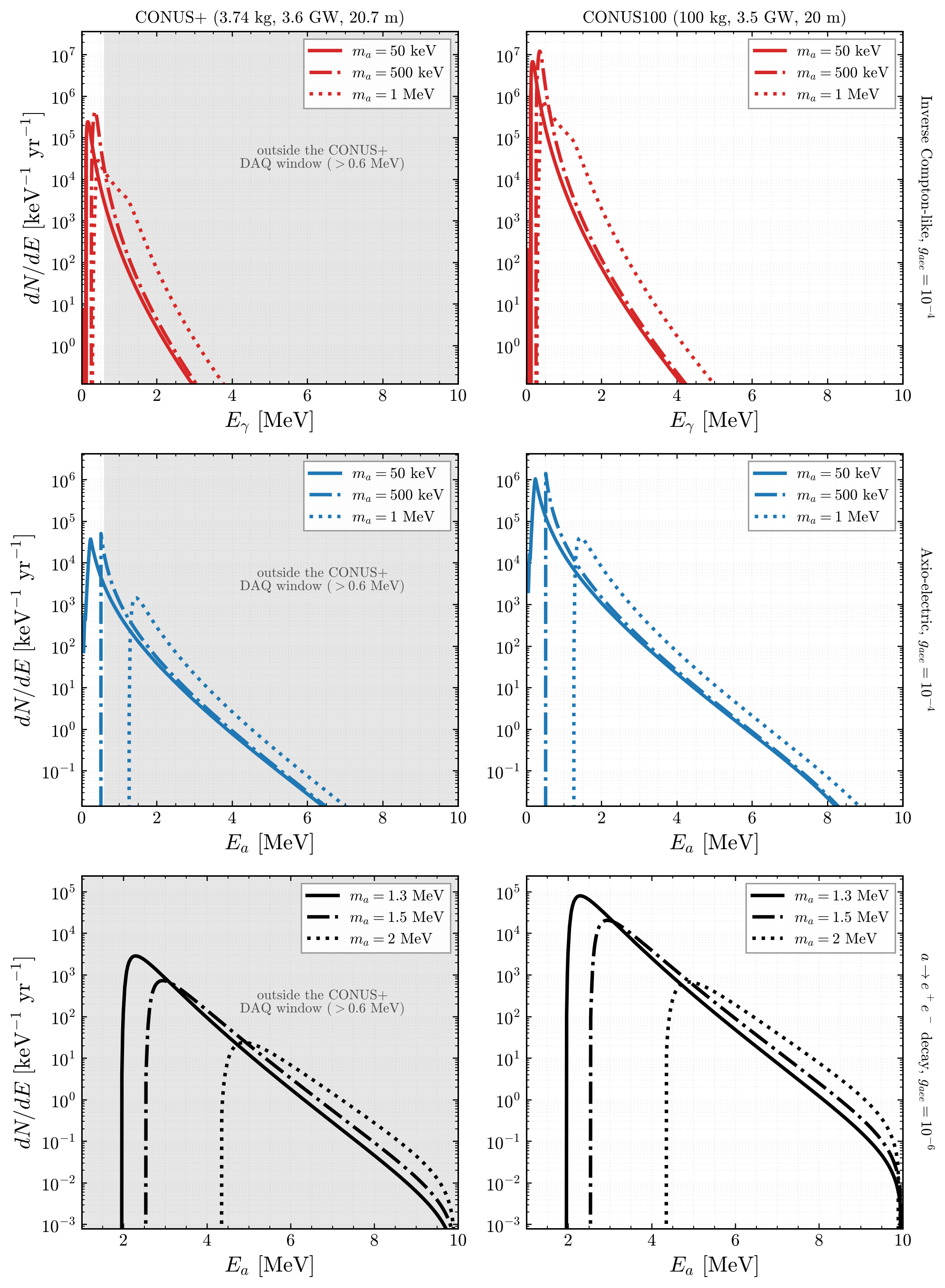}
    \caption{Differential event rates per keV and per full-power year for CONUS+
    (left) and CONUS100 (right): inverse Compton-like scattering (top) and
    axio-electric absorption (middle) for $g_{aee}=10^{-4}$, and
    $a\to e^+e^-$ decays inside the detector (bottom) for $g_{aee}=10^{-6}$;
    the line style gives $m_a$.  No efficiency or energy window is applied;
    the shaded area marks the energies above the CONUS+ DAQ window,
    $0.6~\mathrm{MeV}$.}
    \label{fig:electron-alps_rate}
\end{figure}

Figures~\ref{fig:DP_rates_grid}, \ref{fig:conus_alps} and
\ref{fig:electron-alps_rate} show representative differential event rates in
CONUS+ and CONUS100 for dark photons and for ALPs coupled to photons and to
electrons, respectively. The dark photon rates use the same
production and detection cross sections as the sensitivity of
Sec.~\ref{subsec:dp_discussion}, with the CONUS+ and CONUS100 parameters of
Table~\ref{tab:alp_benchmarks}; they scale as $\epsilon^4$.  The ALP rates are shown before any
energy window is applied: for CONUS+ only the part below $0.6~\mathrm{MeV}$
enters the projections, so that the $a\to e^+e^-$ signal lies entirely
outside its window, whereas for CONUS100 the window is assumed to extend to
$10~\mathrm{MeV}$.  The small structure near $0.1~\mathrm{MeV}$ in the
photon-coupling rates reflects the uranium K edge in the photon attenuation
of the production medium.  The sharp edges of the electron-coupling rates
are kinematic limits of the production and detection processes; for
instance, the inverse Compton-like photon spectrum is bounded from below by
$\simeq m_e/2$, and for $m_a>m_e$ the Compton-like production gives
$E_a\gtrsim(m_a^2+m_e^2)/(2m_e)$.

\subsection{Projected sensitivities}
\label{Discussion}

The sensitivities presented below are obtained from the signal predictions
derived in the previous sections. Different confidence-level conventions
are retained for the two scenarios considered in this work: the dark photon projection is quoted at $95\%$ C.L., following the convention adopted in Ref.~\cite{Park:2017}, whereas the ALP projections are presented at $90\%$ C.L., consistently with Ref.~\cite{AristizabalSierra:2020rom}.

\subsubsection{Dark photon}
\label{subsec:dp_discussion}

\begin{figure}
    \centering
    \includegraphics[width=1\linewidth]{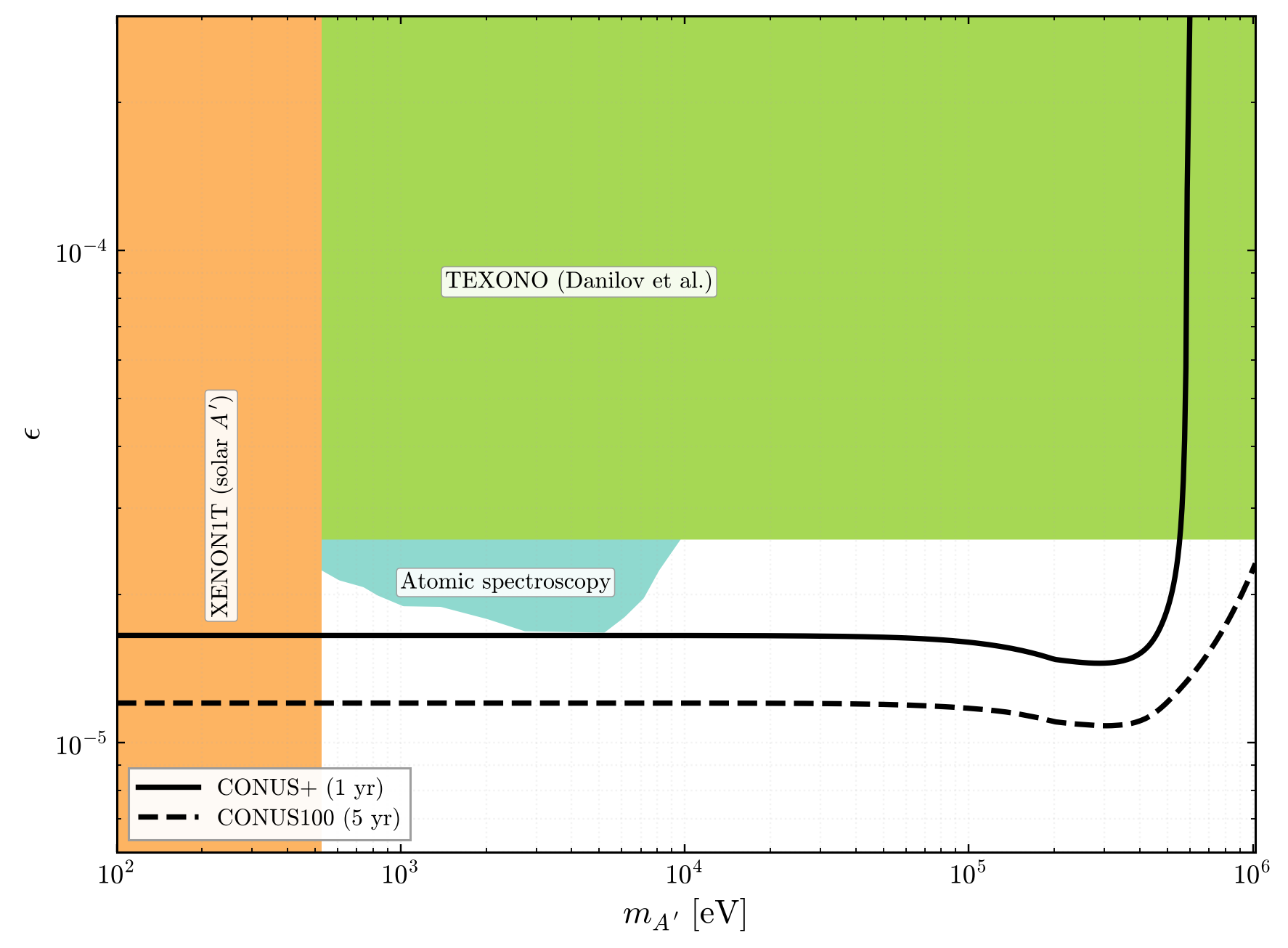}
    \caption{Projected $95\%$ C.L. sensitivity to $\epsilon$ for CONUS+ (solid) and
    CONUS100 (dashed), with $b=10$~DRU and unit efficiency between
    $0.2$ and $0.6~\mathrm{MeV}$ (CONUS+, present DAQ) or $10~\mathrm{MeV}$
    (CONUS100, assumed).  Shaded regions are laboratory limits: TEXONO as
    re-analysed in Ref.~\cite{Danilov:2018tqg} (green), atomic
    spectroscopy~\cite{Jaeckel:2010xx} (aqua) and solar dark photons in
    XENON1T~\cite{An:2020bxd} (orange), from Ref.~\cite{AxionLimits}.}
    \label{fig:darkphoton_sensitivity}
\end{figure}

The projected sensitivity to the kinetic-mixing parameter $\epsilon$ is
obtained from the signal prediction of
Eq.~\eqref{eq:dp_signal_events}. We assume a flat benchmark background
of $b=10$~DRU ($1~{\rm DRU}=1~{\rm event\,kg^{-1}\,day^{-1}\,keV^{-1}}$)
within the accepted deposited energy interval, $0.2$--$0.6~\mathrm{MeV}$ for CONUS+ (present DAQ) and $0.2$--$10~\mathrm{MeV}$ for CONUS100 (assumed extended readout), with the same upper edges as for the ALPs; the lower edge is the one adopted for the photon spectrum in Ref.~\cite{Park:2017}.  The expected number of
background events is therefore $B=b\,M_{\rm det}\,\Delta t\,\Delta E$, where $M_{\rm det}$ is the active detector mass, $\Delta t$ the exposure time and $\Delta E$ the width of the accepted interval.

Following the counting prescription of
Ref.~\cite{Park:2017}, the expected $95\%$ C.L. upper limit on
a background-only signal is approximated by
\begin{equation}
    S_{95}=1.64\sqrt{B},
    \label{eq:dp_s95}
\end{equation}
the one-sided $95\%$ quantile appropriate for an upper limit; Ref.~\cite{Park:2017} used the two-sided value $1.96$, as noted in Ref.~\cite{Danilov:2018tqg}.
Using the $\epsilon^4$ dependence of the total event yield,
Eq.~\eqref{eq:dp_epsilon_scaling}, the corresponding sensitivity is
\begin{equation}
    \epsilon_{95}(m_{A'})
    =
    \left[
        \frac{S_{95}}
             {N_0(m_{A'})}
    \right]^{1/4},
    \label{eq:dp_epsilon95}
\end{equation}
with $N_0(m_{A'})$ from Eq.~\eqref{eq:dp_epsilon_scaling}.

The resulting sensitivities are shown in Fig.~\ref{fig:darkphoton_sensitivity} over the interval $100~{\rm eV}\leq m_{A'}\leq2m_e$.  At low masses $N_0$ depends only weakly on $m_{A'}$, and the sensitivity forms a plateau at $\epsilon\simeq1.65\times10^{-5}$ for CONUS+ and $1.2\times10^{-5}$ for CONUS100.  At $m_{A'}\sim0.1$--$0.3~\mathrm{MeV}$ the finite-mass kinematics redistributes the dark photon spectrum within the accepted window, and the sensitivity improves slightly, to $1.45\times10^{-5}$ and $1.08\times10^{-5}$ at $m_{A'}\simeq0.3~\mathrm{MeV}$.

For $m_{A'}>E_{\rm dep}^{\rm min}$, the available integration interval
progressively decreases because its lower boundary rises with $m_{A'}$. For CONUS+ the corresponding loss of accepted phase space eventually
dominates the mass dependence of the production and detection rates,
causing the sensitivity to deteriorate rapidly. In the limit $m_{A'}\rightarrow E_{\rm dep}^{\rm max}=0.6~{\rm MeV}$
the accepted interval collapses,
$N_0(m_{A'})\rightarrow0$, and the projected upper limit on
$\epsilon$ diverges, so the CONUS+ curve terminates at this endpoint.  With
its readout extended to $10~\mathrm{MeV}$, CONUS100 keeps its sensitivity up
to the $A'\to e^+e^-$ threshold, reaching $\epsilon\simeq2.3\times10^{-5}$ at
$m_{A'}=1~\mathrm{MeV}$.

The relative improvement obtained with CONUS100 can also be understood
from the scaling of the signal and background. In the
background-dominated regime,
\begin{equation}
    \epsilon_{95}
    \propto
    (M_{\rm det}\Delta t)^{-1/8}
    \left(\frac{L^2}{P}\right)^{1/4},
    \label{eq:dp_sensitivity_scaling}
\end{equation}
for fixed background density and accepted energy interval. The larger
target mass and exposure of the CONUS100 benchmark alone would lead to an
improvement of approximately a factor of $1.9$ in the low-mass
sensitivity relative to CONUS+. Since most of the signal lies below
$1~\mathrm{MeV}$ while the background grows with the width of the window,
$\epsilon_{95}\propto(\Delta E)^{1/8}$, the wider CONUS100 window reduces this
factor to $1.4$ at low mass, the price of its larger mass reach.  A
dedicated region of interest of $0.2$--$1.0~\mathrm{MeV}$ would give
$\epsilon\simeq(0.8$--$0.9)\times10^{-5}$ for $m_{A'}\lesssim0.5~\mathrm{MeV}$.

For comparison, Fig.~\ref{fig:darkphoton_sensitivity} displays the
laboratory limits in this mass range~\cite{AxionLimits}: the TEXONO reactor
search as re-analysed by Danilov et al.~\cite{Danilov:2018tqg},
$\epsilon<2.6\times10^{-5}$ for $m_{A'}\gtrsim20~\mathrm{eV}$, and atomic
spectroscopy~\cite{Jaeckel:2010xx}, which reaches $1.7\times10^{-5}$ near
$5~\mathrm{keV}$.  As for the ALPs, we also show the search for solar dark
photons absorbed in XENON1T~\cite{An:2020bxd}, which excludes
$\epsilon\gtrsim10^{-14}$--$10^{-12}$ up to $m_{A'}\simeq0.5~\mathrm{keV}$, the
largest mass for which the solar flux was computed there.  Searches that
assume that the dark photon is the dark matter are not shown. Including the secondary photons of Ref.~\cite{Du:2023tqs} would strengthen the TEXONO limit by at most $\sim30\%$~\cite{Demidov:2026cmt}.  Stellar and other astrophysical
exclusions are not included.  For $0.5~\mathrm{keV}\lesssim m_{A'}\lesssim0.5~\mathrm{MeV}$, CONUS+ lies below the TEXONO limit by a factor
$\simeq1.6$ and CONUS100 by $\simeq2.2$ in $\epsilon$.  As its window closes, CONUS+ crosses the TEXONO limit at
$m_{A'}\simeq0.56~\mathrm{MeV}$, whereas CONUS100 stays below it up to $2m_e$; near $5~\mathrm{keV}$
the CONUS+ projection touches the spectroscopy bound.

\subsubsection{Axion-like particles}
\label{subsec:alp_discussion}

The ALP signal yields derived in Sec.~\ref{Model} are translated into projected sensitivities with the same flat background benchmark as for the dark photon, $b=10$~DRU, for all detection channels.  This value is a phenomenological benchmark and should not be interpreted as a measurement of the CONUS+ background over the full energy range relevant to the different channels.  For a continuum signal the expected background is $B_X=b\,M_{\rm det}\,\Delta t\,\Delta E_X$, where $\Delta E_X$ is the full accepted energy interval of the corresponding signal; Ref.~\cite{AristizabalSierra:2020rom} does not specify the interval used for the background. For the monochromatic signatures of
the magnetic-transition analysis (inverse Primakoff conversion,
axio-electric absorption and in-detector decays of the
$2.230~\mathrm{MeV}$ ALPs), the DRU density is converted to a single-bin
background using a $1~\mathrm{keV}$ analysis bin. This choice is a convention of the present projection and should not be
interpreted as a detector resolution or as an experimentally optimized
region of interest.  For a germanium resolution of about $2~\mathrm{keV}$
(FWHM) at $2.23~\mathrm{MeV}$, a bin containing essentially the whole line
would be $\simeq3~\mathrm{keV}$ wide, which would weaken the projected reach
on the coupling products by about $30\%$.  The escape of annihilation
photons after $a\to e^+e^-$ decays, which would move part of the signal out
of the full-energy line, is not modelled.

\begin{figure}
    \centering
    \includegraphics[width=1\linewidth]{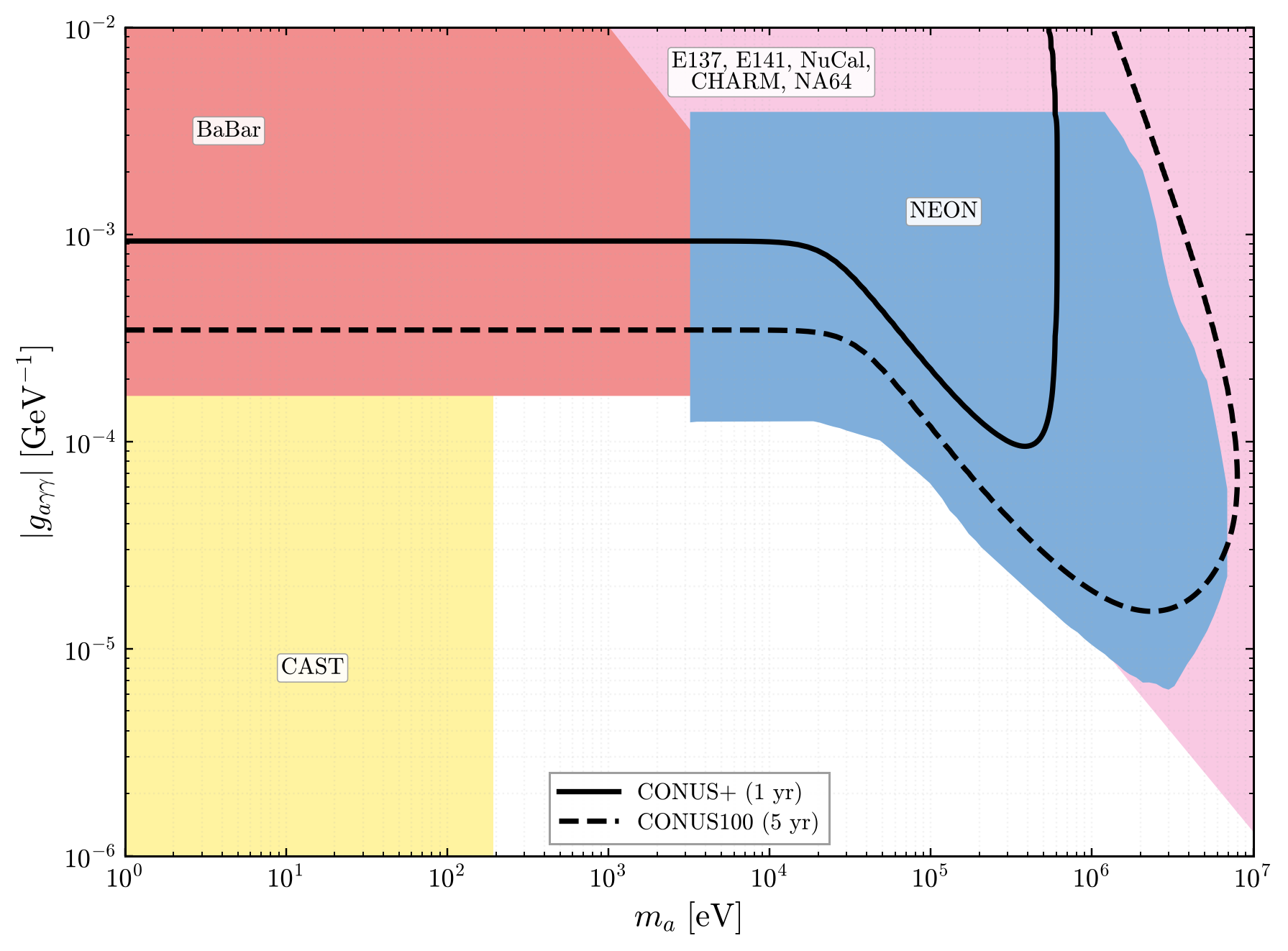}
    \caption{Projected $90\%$ C.L. sensitivity to $|g_{a\gamma\gamma}|$ for CONUS+
    (solid) and CONUS100 (dashed), with $b=10$~DRU and unit efficiency between
    $25~\mathrm{keV}$ and $0.6~\mathrm{MeV}$ (CONUS+, present DAQ) or
    $10~\mathrm{MeV}$ (CONUS100, assumed).  Shaded regions are experimental
    limits: CAST~\cite{CAST:2007jps,CAST:2017uph} (yellow),
    BaBar~\cite{Dolan:2017osp} (red), the beam dumps
    E137~\cite{Bjorken:1988as,Dolan:2017osp}, E141~\cite{Riordan:1987aw},
    NuCal~\cite{Blumlein:1990ay}, CHARM~\cite{CHARM:1985anu} and
    NA64~\cite{NA64:2020qwq}, as combined in Ref.~\cite{AxionLimits} (pink),
    and NEON~\cite{NEON:2024kwv} (blue, $95\%$ C.L.).}
    \label{fig:gagg_sensitivity}
\end{figure}

\begin{figure}
    \centering
    \includegraphics[width=1\linewidth]{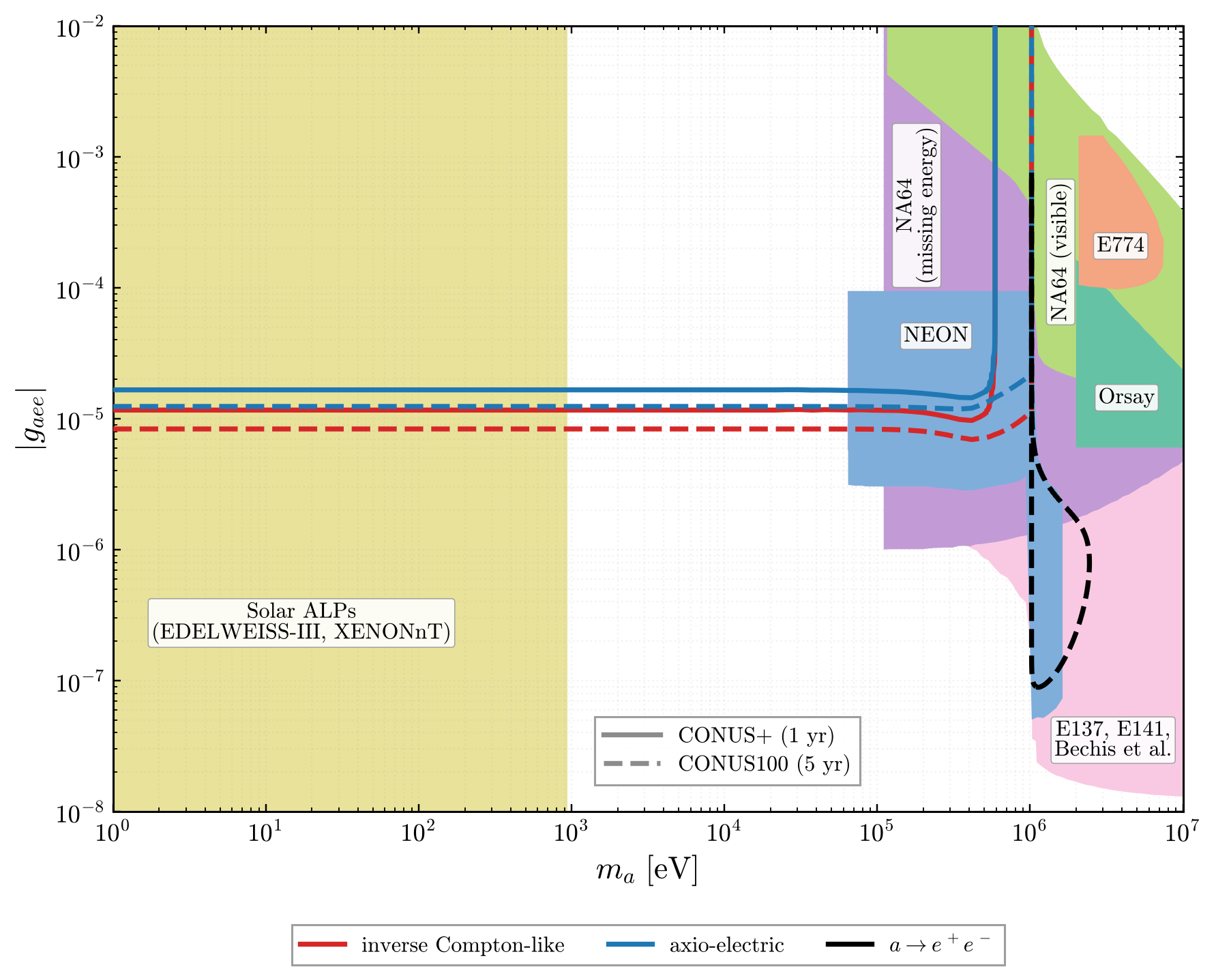}
    \caption{Projected $90\%$ C.L. sensitivity to $|g_{aee}|$ for CONUS+ (solid) and
    CONUS100 (dashed), with $b=10$~DRU and efficiency = 1 between
    $25~\mathrm{keV}$ and $0.6~\mathrm{MeV}$ (CONUS+, present DAQ) or
    $10~\mathrm{MeV}$ (CONUS100, assumed).  Shaded regions are experimental
    limits: solar-ALP searches with EDELWEISS-III~\cite{EDELWEISS:2018tde}
    and XENONnT~\cite{XENON:2022ltv} (khaki); the electron beam dumps
    E137~\cite{Bjorken:1988as}, E141~\cite{Riordan:1987aw} and Bechis et
    al.~\cite{Bechis:1979kp} (pink, one region as in
    Ref.~\cite{NEON:2024kwv}); E774~\cite{Bross:1989mp} (orange) and
    Orsay~\cite{Davier:1989wz} (teal), from Ref.~\cite{NA64:2021aiq}; NA64
    missing-energy (purple) and visible-decay (green) searches, from
    Ref.~\cite{Eberhart:2025lep}; and NEON~\cite{NEON:2024kwv} (blue,
    $95\%$ C.L.).}
    \label{fig:gaee_sensitivity}
\end{figure}

Following Ref.~\cite{AristizabalSierra:2020rom}, the projected $90\%$ C.L.
sensitivity is determined from
\begin{equation}
    \chi^2
    =
    \frac{S^2}{S+B},
    \qquad
    \Delta\chi^2=2.71,
    \label{eq:alp_chi2_discussion}
\end{equation}
where $S$ and $B$ denote the expected signal and background event counts,
respectively.  Throughout the ALP projections we assume unit signal
efficiency inside the accepted energy window of
Table~\ref{tab:alp_benchmarks}: $25~\mathrm{keV}$--$0.6~\mathrm{MeV}$ for
CONUS+, set by its present DAQ~\cite{ConusPlus:2024}, and
$25~\mathrm{keV}$--$10~\mathrm{MeV}$ for CONUS100, which assumes an
extended readout, and the full-power-equivalent exposures of
Table~\ref{tab:alp_benchmarks}. For the ALP masses and couplings
considered here the ALP energy is fully deposited in the germanium,
$E_{\rm dep}=E_a$, so for CONUS+ only ALPs with $E_a\le0.6~\mathrm{MeV}$
contribute, and the continuum background $B_X$ is integrated only up to
$0.6~\mathrm{MeV}$.  Two signals are therefore outside the CONUS+ window: the
$a\to e^+e^-$ decays, which deposit at least $2m_e\simeq1.022~\mathrm{MeV}$,
and all signals from the $2.230~\mathrm{MeV}$ $p(n,\gamma)d$ line.  CONUS+ thus
appears only in the single-coupling projections of
Figs.~\ref{fig:gagg_sensitivity} and \ref{fig:gaee_sensitivity}, and
the nuclear-transition and $a\to e^+e^-$ results are shown for CONUS100 alone. Systematic uncertainties and nuisance
parameters are not included, so the resulting curves should be interpreted
as idealized sensitivity projections rather than detector-level likelihood
results.

\begin{figure}
    \centering
    \includegraphics[width=1\linewidth]{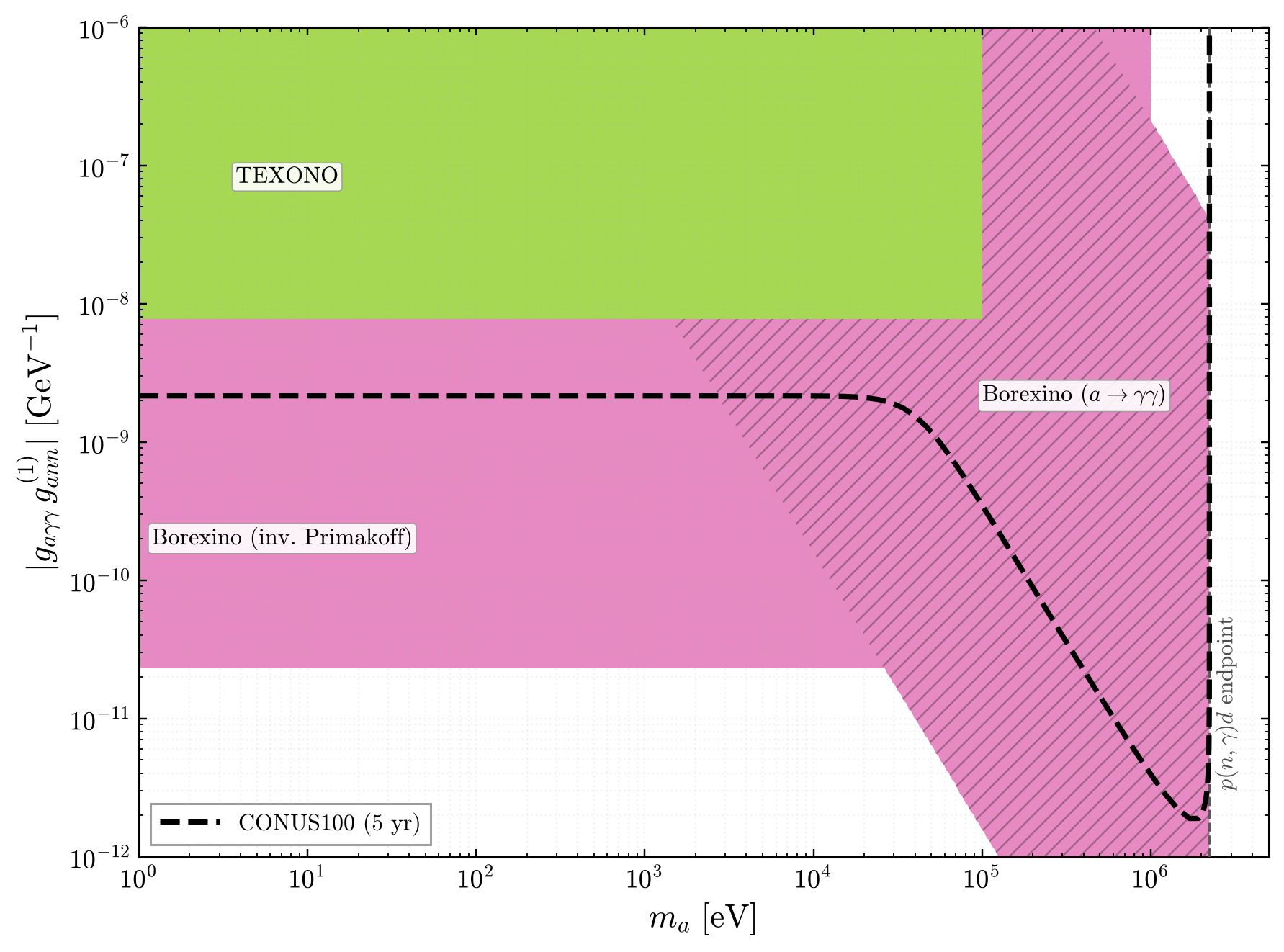}
    \caption{Projected $90\%$ C.L. sensitivity to $|g_{a\gamma\gamma}g_{ann}^{(1)}|$
    for ALPs from the $2.230~\mathrm{MeV}$ $p(n,\gamma)d$ line, for CONUS100
    (dashed) with $b=10$~DRU in a $1~\mathrm{keV}$ bin, unit efficiency up to
    $10~\mathrm{MeV}$ (assumed) and $g_{ann}^{(1)}\le1$; the line is outside
    the $0.6~\mathrm{MeV}$ DAQ window of CONUS+.  Shaded regions are
    experimental limits: TEXONO~\cite{TEXONO:2006axion} (green) and
    Borexino~\cite{Borexino:2025} from inverse Primakoff conversion (magenta)
    and from $a\to\gamma\gamma$ decays of solar ALPs (magenta, hatched;
    $g_{ann}^{(1)}\le1$).}
    \label{fig:gagg_gann_sensitivity}
\end{figure}

\begin{figure}
    \centering
    \includegraphics[width=1\linewidth]{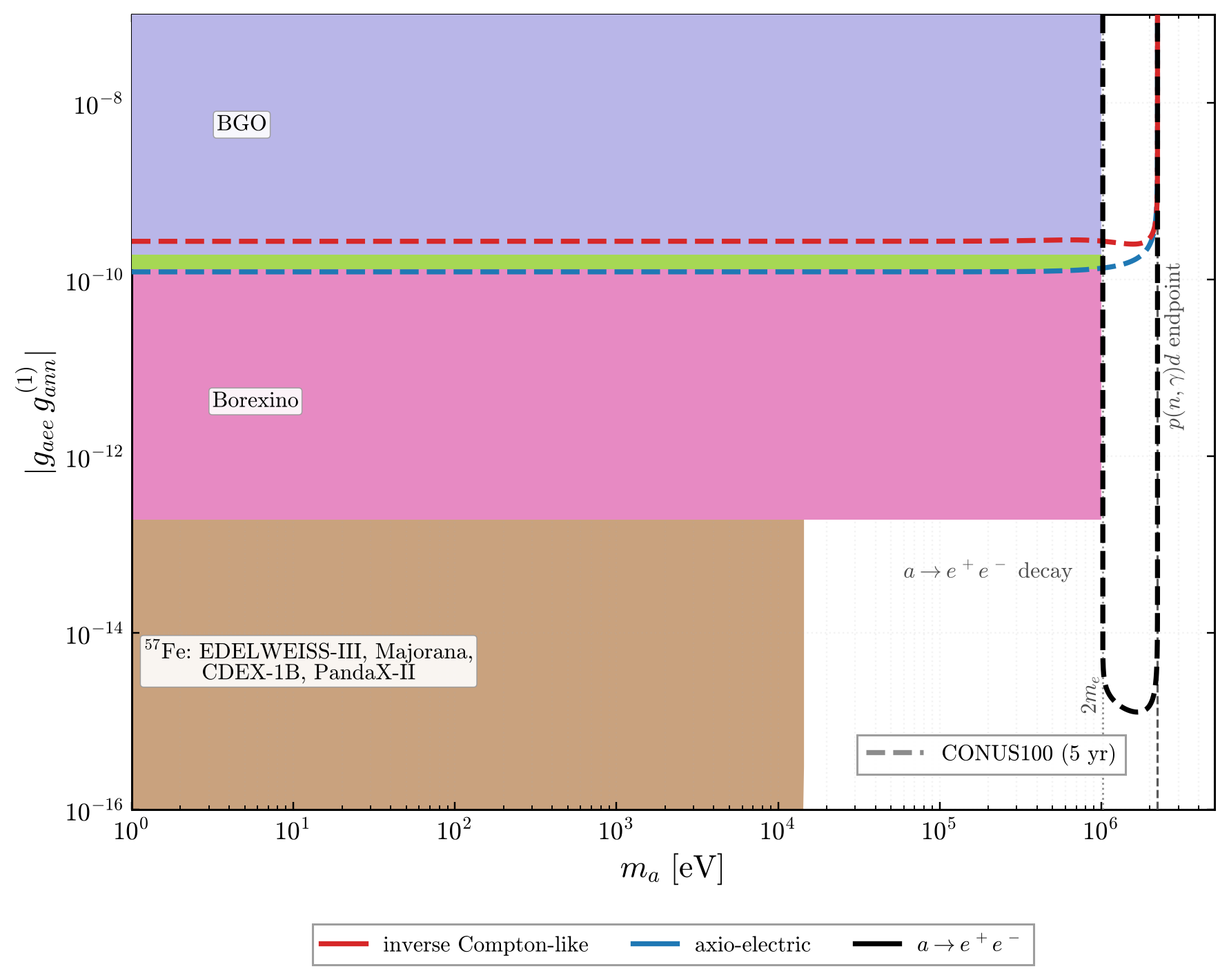}
    \caption{Projected $90\%$ C.L. sensitivity to $|g_{aee}g_{ann}^{(1)}|$ for ALPs
    from the $2.230~\mathrm{MeV}$ $p(n,\gamma)d$ line, for CONUS100 (dashed)
    with $b=10$~DRU in a $1~\mathrm{keV}$ bin and efficiency 1 up to
    $10~\mathrm{MeV}$ (assumed); the line is outside the $0.6~\mathrm{MeV}$
    DAQ window of CONUS+.  Shaded regions are experimental limits:
    BGO~\cite{Derbin:2014xzr} (lavender), Borexino~\cite{Borexino:2025}
    (magenta), TEXONO~\cite{TEXONO:2006axion} (thin light-green band just
    above the blue contour) and the $^{57}$Fe solar-ALP
    searches~\cite{EDELWEISS:2018tde,Majorana:2016hsl,CDEX:2019exx,PandaX:2017ock} (brown).}
    \label{fig:gaee_gann_sensitivity}
\end{figure}

\begin{figure}
    \centering
    \includegraphics[width=1\linewidth]{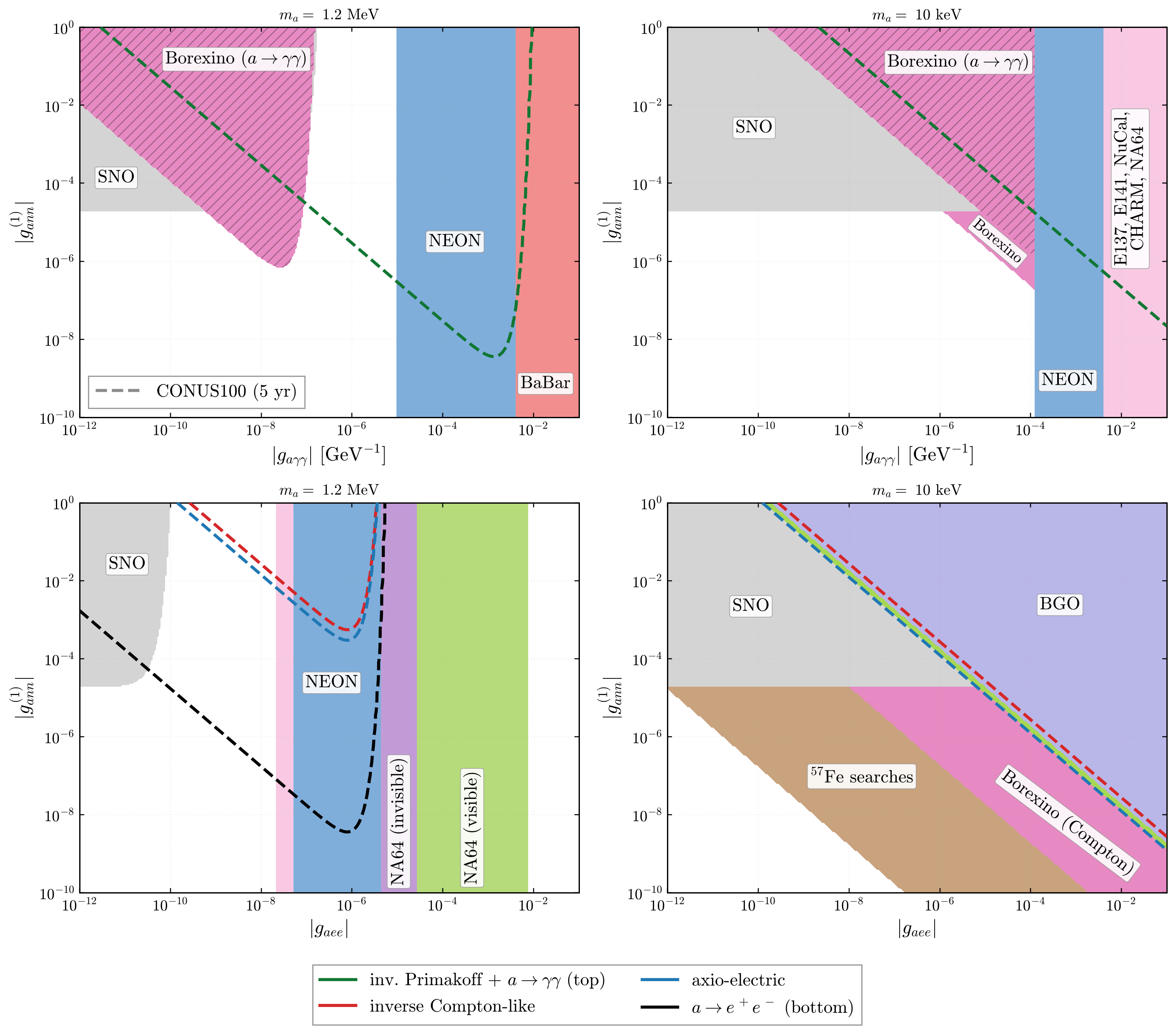}
    \caption{Projected $90\%$ C.L. sensitivities of CONUS100 (dashed) in the
    $(g_{a\gamma\gamma},g_{ann}^{(1)})$ (top) and $(g_{aee},g_{ann}^{(1)})$
    (bottom) planes at $m_a=1.2~\mathrm{MeV}$ (left) and $10~\mathrm{keV}$
    (right), with $b=10$~DRU in a $1~\mathrm{keV}$ bin and efficiency assumed to be 1 up to
    $10~\mathrm{MeV}$ (assumed).  CONUS+ is not shown, since the
    $2.230~\mathrm{MeV}$ line lies outside its $0.6~\mathrm{MeV}$ DAQ window.
    Shaded regions are experimental limits:
    SNO~\cite{Bhusal:2020bvx} (grey); on $g_{a\gamma\gamma}$,
    BaBar~\cite{Dolan:2017osp} (red), the beam dumps E137, E141, NuCal, CHARM
    and NA64~\cite{Bjorken:1988as,Riordan:1987aw,Blumlein:1990ay,CHARM:1985anu,NA64:2020qwq,AxionLimits}
    (pink) and NEON~\cite{NEON:2024kwv} (blue); on $g_{aee}$, the electron
    beam dumps E137, E141 and Bechis et al.~\cite{Bjorken:1988as,Riordan:1987aw,Bechis:1979kp,Eberhart:2025lep}
    (pink), NEON (blue) and the NA64 missing-energy (purple) and
    visible-decay (green) searches~\cite{Eberhart:2025lep,NEON:2024kwv}; and
    the product bounds from TEXONO~\cite{TEXONO:2006axion} (light green),
    Borexino~\cite{Borexino:2025} (magenta; hatched: $a\to\gamma\gamma$
    decays of solar ALPs), BGO~\cite{Derbin:2014xzr} (lavender) and the
    $^{57}$Fe searches~\cite{EDELWEISS:2018tde,Majorana:2016hsl,CDEX:2019exx,PandaX:2017ock} (brown).}
    \label{fig:coupling_planes}
\end{figure}

In Figs.~\ref{fig:gagg_sensitivity}--\ref{fig:coupling_planes} the
projections are compared only with current \emph{experimental} limits,
i.e.\ searches with a laboratory source (reactors, accelerators, beam dumps)
or with a terrestrial detector exposed to the solar ALP flux (helioscopes,
Borexino, BGO, SNO and $^{57}$Fe searches).  Bounds from stellar cooling,
supernovae and cosmology, as well as searches that assume that the ALP
constitutes the galactic dark matter, are not shown.  All limits are at
$90\%$ C.L., except the NEON reactor limits~\cite{NEON:2024kwv} and the SNO
bound~\cite{Bhusal:2020bvx}, which are quoted at $95\%$ C.L.  In all
figures, solid (dashed) curves correspond to CONUS+ (CONUS100), as indicated
in a box inside each plot; when several detection channels are shown, their
colours are given in the legend below the plot.

\paragraph{\textbf{Photon--ALP coupling}.}

Figure~\ref{fig:gagg_sensitivity} shows the projected sensitivity to
$g_{a\gamma\gamma}$ when the same coupling controls both reactor production
and detection.  At low ALP masses, both CONUS+ and CONUS100 exhibit a
mass-independent sensitivity, $|g_{a\gamma\gamma}|\simeq 9.6\times10^{-4}$
($3.5\times10^{-4}$)~$\mathrm{GeV}^{-1}$ for CONUS+ (CONUS100).  This
corresponds to the long-lived regime, where Primakoff production followed by
inverse Primakoff scattering dominates and varies only weakly with $m_a$.
For $m_a\gtrsim50~\mathrm{keV}$ the $a\to\gamma\gamma$ decay inside the
germanium becomes the dominant channel and the sensitivity improves
approximately as $m_a^{-2}$, reaching $1.5\times10^{-5}~\mathrm{GeV}^{-1}$
at $m_a\simeq2~\mathrm{MeV}$ for CONUS100.  For CONUS+ the improvement stops
at $|g_{a\gamma\gamma}|\simeq1.0\times10^{-4}~\mathrm{GeV}^{-1}$ for
$m_a\simeq0.35~\mathrm{MeV}$, and its contour closes at
$m_a\simeq0.6~\mathrm{MeV}$, since heavier ALPs carry
$E_a\ge m_a$ above the DAQ range.  At
still larger masses and couplings, the reduction of the produced flux
together with the decay of the ALPs before they reach the detector closes
the sensitive region.  The relative improvement of CONUS100 is not uniform
in mass, since the signal composition changes between the scattering- and
decay-dominated regimes.

The relevant experimental constraints in this mass range are set by the
BaBar mono-photon search~\cite{Dolan:2017osp} for
$m_a\lesssim60~\mathrm{keV}$
($|g_{a\gamma\gamma}|<1.7\times10^{-4}~\mathrm{GeV}^{-1}$) and, above, by
beam-dump experiments, E137~\cite{Bjorken:1988as,Dolan:2017osp}, E141~\cite{Riordan:1987aw}, NuCal~\cite{Blumlein:1990ay}, CHARM~\cite{CHARM:1985anu} and NA64~\cite{NA64:2020qwq}, as combined in
Ref.~\cite{AxionLimits}; the helioscope CAST~\cite{CAST:2007jps,CAST:2017uph}
dominates for $m_a\lesssim200~\mathrm{eV}$.  Two further bounds are not
drawn because they add no visible region in this window:
NOMAD~\cite{NOMAD:2000usb} ($m_a\lesssim0.2~\mathrm{keV}$) lies within $6\%$
of BaBar, and the recast of MiniBooNE beam-dump data~\cite{Capozzi:2023ffu}
lies within $5\%$ of the beam-dump bound below $10~\mathrm{MeV}$.  Between
$3~\mathrm{keV}$ and $\sim7~\mathrm{MeV}$ the reactor search with the NEON
NaI(Tl) detector~\cite{NEON:2024kwv} adds to these constraints; NEON is the
strongest bound for $3~\mathrm{keV}\lesssim m_a\lesssim1~\mathrm{MeV}$,
reaching $|g_{a\gamma\gamma}|\simeq1.3\times10^{-4}~\mathrm{GeV}^{-1}$ at
$10~\mathrm{keV}$ and $9\times10^{-6}~\mathrm{GeV}^{-1}$ at
$1.2~\mathrm{MeV}$.  With $b=10$~DRU the projected CONUS reach lies inside
the experimentally excluded region over the whole mass range; the closest
approach is at $m_a\simeq0.5~\mathrm{MeV}$, where the CONUS100
projection is a factor $\simeq1.8$ above the NEON limit, while CONUS+
remains at least a factor $\simeq3.5$ above it (at $m_a\simeq0.1~\mathrm{MeV}$).

\paragraph{\textbf{Electron--ALP coupling}.}

The corresponding single-coupling sensitivity to $g_{aee}$ is shown in
Fig.~\ref{fig:gaee_sensitivity}.  Below the $e^+e^-$ threshold the inverse
Compton-like and axio-electric yields scale as $g_{aee}^4$ and the
sensitivities are nearly mass independent, resulting in approximately
horizontal contours: $|g_{aee}|\simeq 1.2\times10^{-5}$ ($8.5\times10^{-6}$)
for inverse Compton-like scattering and $1.7$$\times10^{-5}$
($1.2\times10^{-5}$) for axio-electric absorption at CONUS+ (CONUS100).
The CONUS+ contours end at $m_a\simeq0.6~\mathrm{MeV}$, the upper end of its
DAQ window.  The smaller window also lowers the CONUS+ background, which is
why its scattering reach is slightly better than with a window up to
$10~\mathrm{MeV}$.

A qualitatively different structure appears once
$m_a>2m_e\simeq1.022~\mathrm{MeV}$, where the decay $a\to e^+e^-$ of
Eq.~\eqref{eq:decay_width_gaee} becomes kinematically accessible.  The
finite lifetime then modifies both the flux reaching the detector and the
probability for an ALP to decay inside the germanium, and produces for
CONUS100 a closed
decay region between $m_a=2m_e$ and $\simeq2.4$~MeV, extending down
to $|g_{aee}|\simeq9\times10^{-8}$; CONUS+ has no such region, since the
$e^+e^-$ pair deposits more than its $0.6~\mathrm{MeV}$ DAQ range.  At small
$g_{aee}$, too few ALPs decay inside the detector; at large $g_{aee}$, a
large fraction of the produced particles decay before reaching it.

Among experimental bounds, solar-ALP searches with XENONnT~\cite{XENON:2022ltv} and
EDELWEISS-III~\cite{EDELWEISS:2018tde}    
exclude $|g_{aee}|\gtrsim2\times10^{-12}$ for $m_a\lesssim1~\mathrm{keV}$.
Above $m_a\simeq60~\mathrm{keV}$ the laboratory bounds are taken from the
reanalysis of leptophilic ALPs in Ref.~\cite{Eberhart:2025lep} (the E137
electron beam dump and the NA64 missing-energy and visible-decay searches)
and from Ref.~\cite{NEON:2024kwv} (the NEON reactor search, NA64, and the
beam dumps E137~\cite{Bjorken:1988as}, E141~\cite{Riordan:1987aw} and
Bechis et al.~\cite{Bechis:1979kp}).  The E774~\cite{Bross:1989mp} and
Orsay~\cite{Davier:1989wz} electron beam dumps are taken from
Ref.~\cite{NA64:2021aiq}, with $g_{aee}=e\,\epsilon$ in terms of the coupling
$\epsilon$ used there; the Orsay region is drawn down to the lower end of
the figure of that reference, $|g_{aee}|\simeq6\times10^{-6}$, below which
E137 already excludes.  The NA64 missing-energy search excludes
$|g_{aee}|\gtrsim(1\text{--}2.5)\times10^{-6}$ for
$0.1~\mathrm{MeV}\lesssim m_a\lesssim2m_e$, NEON reaches
$|g_{aee}|\simeq3\times10^{-6}$ between $60$ and $900~\mathrm{keV}$, and
above $2m_e$ the beam dumps exclude
$2\times10^{-8}\lesssim|g_{aee}|\lesssim10^{-4}$.  The CONUS scattering
contours above $60~\mathrm{keV}$ and the whole $a\to e^+e^-$ region
therefore lie inside the excluded region.  For
$1~\mathrm{keV}\lesssim m_a\lesssim60~\mathrm{keV}$ none of the displayed
limits covers the CONUS contours.  This window reflects the mass ranges
shown in Refs.~\cite{NEON:2024kwv,Eberhart:2025lep}: the NA64
missing-energy bound does not depend on $m_a$ for $m_a\ll m_e$ and is
expected to extend to lower masses, so the window should not be read as a
genuine sensitivity gain.

\paragraph{\textbf{Nuclear production with photon detection}.}

Figure~\ref{fig:gagg_gann_sensitivity} shows the projected sensitivity to
the product $|g_{a\gamma\gamma}g_{ann}^{(1)}|$ for the isovector benchmark
of Eq.~\eqref{eq:gann_isovector_benchmark} and the dominant
$2.230~\mathrm{MeV}$ $p(n,\gamma)d$ line.  Detection proceeds through
inverse Primakoff conversion and $a\to\gamma\gamma$ decay.  The coupling
product is used as the presentation variable, while the separate dependence
of propagation and decay on $g_{a\gamma\gamma}$ is retained in the
calculation: for each mass, the curve gives the smallest product for which
some pair $(g_{a\gamma\gamma},g_{ann}^{(1)})$ with $g_{ann}^{(1)}\le1$
reaches the $90\%$ C.L. threshold.

Only CONUS100 is shown: the $2.230~\mathrm{MeV}$ signal lies above the
$0.6~\mathrm{MeV}$ DAQ range of CONUS+.
At low masses the sensitivity is governed by the long-lived scattering
contribution, $|g_{a\gamma\gamma}g_{ann}^{(1)}|\simeq2.2\times10^{-9}~\mathrm{GeV}^{-1}$
for CONUS100.  At larger
masses, $a\to\gamma\gamma$ decays take over and the reach improves to
$1.9\times10^{-12}~\mathrm{GeV}^{-1}$ at
$m_a\simeq1.7~\mathrm{MeV}$.  As $m_a$ approaches the
$2.230~\mathrm{MeV}$ transition energy, the production phase space closes as
$p_a^3$ and the minimal product diverges; the sharp upward turn near the
endpoint is therefore determined by the source kinematics rather than by
the detector response.

The same isovector combination is constrained by TEXONO and by searches for
the $5.5~\mathrm{MeV}$ solar ALPs from $p(d,{}^3\mathrm{He})A$, whose
emission is controlled by $g_{3AN}\equiv g_{ann}^{(1)}$.
TEXONO~\cite{TEXONO:2006axion} obtained
$|g_{a\gamma\gamma}g_{ann}^{(1)}|<7.7\times10^{-9}~\mathrm{GeV}^{-1}$ for
$m_a<10^5~\mathrm{eV}$.  The complete Borexino data set~\cite{Borexino:2025}
gives $|g_{A\gamma}g_{3AN}|\le2.3\times10^{-11}~\mathrm{GeV}^{-1}$ from
inverse Primakoff conversion ($m_A<1~\mathrm{MeV}$) and, from
$A\to\gamma\gamma$ decays inside the detector,
$|g_{A\gamma}g_{3AN}|\,m_A^2\le1.6\times10^{-2}~\mathrm{GeV^{-1}\,eV^2}$ for
ALPs that reach the Earth.  We draw the latter for $g_{3AN}\le1$, including
the $(p_A/p_\gamma)^3$ production factor and the Sun--Earth survival
probability.  The smallest products accessible to CONUS correspond to
long-lived ALPs and are therefore already excluded by Borexino.  The product
representation, however, hides the region in which the ALP decays between
the Sun and the Earth but survives the $20~\mathrm{m}$ reactor baseline;
this region is discussed below with the coupling planes of
Fig.~\ref{fig:coupling_planes}.

\paragraph{\textbf{Nuclear production with electron detection}.}

The corresponding sensitivity to $|g_{aee}g_{ann}^{(1)}|$ is displayed in
Fig.~\ref{fig:gaee_gann_sensitivity}, with detection through inverse
Compton-like scattering, axio-electric absorption, and $a\to e^+e^-$ decay.
As in the photon-detection case, the coupling product is used only as the
plotting variable.  For the scattering channels the product fully
determines the yield in the long-lived regime.  For the decay channel we
show the lower branch of the sensitive region, obtained in the limit of
small $g_{aee}$ at fixed product, where $P_{\rm surv}\to1$ and
$P_{\rm dec}\propto g_{aee}^2$; with $g_{ann}^{(1)}\le1$ this requires
$g_{aee}\gtrsim10^{-15}$, deep in the long-lived regime.

As for the photon channels, only CONUS100 is shown.
In the long-lived regime the scattering channels reach
$|g_{aee}g_{ann}^{(1)}|\simeq1.2\times10^{-10}$ through
axio-electric absorption and $2.7\times10^{-10}$ through
inverse Compton-like scattering for CONUS100, comparable to
TEXONO~\cite{TEXONO:2006axion} ($1.3\times10^{-10}$, $m_a<1~\mathrm{MeV}$)
and well above the bounds from the $5.5~\mathrm{MeV}$ solar ALPs:
Borexino~\cite{Borexino:2025}, $|g_{Ae}g_{3AN}|\le1.9\times10^{-13}$
($m_A<1~\mathrm{MeV}$), and BGO~\cite{Derbin:2014xzr},
$|g_{Ae}g_{3AN}|<1.9\times10^{-10}$ (quoted for $m_A\to0$; its kinematic
weakening is of a few per cent up to $1~\mathrm{MeV}$).  Below $14.4~\mathrm{keV}$, the
$^{57}$Fe solar-ALP searches (EDELWEISS-III~\cite{EDELWEISS:2018tde},
Majorana Demonstrator~\cite{Majorana:2016hsl}, CDEX-1B~\cite{CDEX:2019exx},
PandaX-II~\cite{PandaX:2017ock}) constrain
$|g_{aee}g_{aN}^{\rm eff}|\lesssim4\times10^{-17}$, which coincides with
$|g_{aee}g_{ann}^{(1)}|$ for $g_{ann}^{(0)}=0$.

The high-mass structure is controlled by two physical thresholds.  The
$a\to e^+e^-$ channel opens at $m_a=2m_e$, and as $m_a$ approaches the
$2.230~\mathrm{MeV}$ transition energy the production phase space vanishes
and all three detection channels deteriorate sharply.  Between the two
thresholds the decay channel produces a U-shaped sensitivity with minimum
$1.3\times10^{-15}$ at $m_a\simeq1.68~\mathrm{MeV}$
for CONUS100.  The published Borexino and BGO bounds are quoted for
$m_A<1~\mathrm{MeV}$, so this region is not covered by existing experimental
limits.

\paragraph{\textbf{Coupling planes at fixed mass}.}

Because the propagation and decay probabilities depend on the detection
coupling separately, the product plots of
Figs.~\ref{fig:gagg_gann_sensitivity} and \ref{fig:gaee_gann_sensitivity}
show only the smallest product that can be probed.  Following
Ref.~\cite{AristizabalSierra:2020rom}, Fig.~\ref{fig:coupling_planes}
shows the full sensitivity in the $(g_{a\gamma\gamma},g_{ann}^{(1)})$ and
$(g_{aee},g_{ann}^{(1)})$ planes for $m_a=1.2~\mathrm{MeV}$, where decays
dominate, and $m_a=10~\mathrm{keV}$, where scattering dominates.  In
addition to the product bounds discussed above, we show the SNO search for
$5.5~\mathrm{MeV}$ solar ALPs detected through deuteron
dissociation~\cite{Bhusal:2020bvx}, which constrains the isovector coupling
alone, $|g_{aN}^{3}|<2\times10^{-5}~\mathrm{GeV}^{-1}$ ($95\%$ C.L.),
i.e.\ $|g_{ann}^{(1)}|=m_N|g_{aN}^{3}|\lesssim1.9\times10^{-5}$, using the
pseudoscalar form of the interaction given in that reference, together with the single-coupling laboratory and reactor
bounds on $g_{a\gamma\gamma}$ and $g_{aee}$ of
Figs.~\ref{fig:gagg_sensitivity} and \ref{fig:gaee_sensitivity} at the
corresponding mass.  Since all the solar bounds require the ALP to reach
the Earth, they are weighted with the Sun--Earth survival probability
evaluated with the detection coupling.

At $m_a=10~\mathrm{keV}$ the CONUS sensitivities are entirely contained in
the experimentally excluded region in both planes.  At
$m_a=1.2~\mathrm{MeV}$, instead, the solar searches lose sensitivity once
the ALP decays before reaching the Earth, whereas it still survives the
reactor baseline.  CONUS100 then probes a region not covered by current
experimental limits for
$10^{-7}\lesssim|g_{a\gamma\gamma}|/\mathrm{GeV}^{-1}\lesssim9\times10^{-6}$
with $|g_{ann}^{(1)}|\gtrsim3\times10^{-7}$, bounded from above in
$g_{a\gamma\gamma}$ by the NEON and beam-dump limits, and, through
$a\to e^+e^-$ decays, for
$3\times10^{-11}\lesssim|g_{aee}|\lesssim2\times10^{-8}$ with
$|g_{ann}^{(1)}|\gtrsim9\times10^{-8}$, bounded from above in $g_{aee}$ by
the E137 beam dump. CONUS+, with its present $0.6~\mathrm{MeV}$ DAQ
range, does not reach these regions.
These are the regions in which a reactor experiment can improve on existing
experimental bounds; astrophysical constraints, not shown here, are
expected to be relevant in the same region.

Taken together, the ALP projections illustrate the complementarity between
continuous-spectrum and monochromatic reactor production.  The former
enables single-coupling searches in $g_{a\gamma\gamma}$ and $g_{aee}$, while
the latter provides sensitivity to products involving the nucleon coupling,
$|g_{a\gamma\gamma}g_{ann}^{(1)}|$ and $|g_{aee}g_{ann}^{(1)}|$.  The
different structures of the projected curves can be traced to a small number
of physical effects: finite ALP lifetimes, the opening of visible decay
channels, and the closure of the available production phase space near the
corresponding kinematic endpoints.  When compared with current experimental
limits only, the single-coupling and minimal-product projections lie inside
regions that are already excluded, while two regions remain open: the
$a\to e^+e^-$ window of $|g_{aee}g_{ann}^{(1)}|$ between $2m_e$ and
$2.230~\mathrm{MeV}$, and, in the coupling planes at MeV masses, the region
of ALPs that decay between the Sun and the Earth but survive the reactor
baseline.

\section{Conclusions}
\label{conclusions}\label{sec:conclusions}

In this work we have investigated the sensitivity of the CONUS+ experiment
and a future CONUS100 configuration to light weakly coupled bosons produced
in a nuclear reactor. Two representative scenarios were considered: a
kinetically mixed dark photon and axion-like particles coupled to photons,
electrons, and nucleons. The short reactor--detector baseline, intense
reactor photon field, and low-background germanium targets make these
experiments sensitive to production and detection mechanisms that are
complementary to conventional neutrino measurements.

For dark photons, we evaluated Compton-like production and inverse
Compton-like detection while retaining the full finite-mass kinematics.
The calculation also incorporates the total photon interaction cross
section in uranium rather than the Compton-only approximation at
few-hundred-keV energies. The resulting sensitivity is approximately
mass independent in the low-mass regime, exhibits a moderate finite-mass
variation at intermediate masses, and rapidly deteriorates as
$m_{A'}$ approaches the upper boundary of the accepted
deposited-energy window, $0.6~{\rm MeV}$ for CONUS+, while CONUS100, with
a readout assumed to extend to $10~{\rm MeV}$, remains sensitive up to
$2m_e$.  For $0.5~{\rm keV}\lesssim m_{A'}\lesssim0.5~{\rm MeV}$, above the reach of solar dark photon searches, the projections
reach $\epsilon\simeq1.65\times10^{-5}$ (CONUS+) and $1.2\times10^{-5}$
(CONUS100), below the TEXONO limit as re-analysed in Ref.~\cite{Danilov:2018tqg}
by factors of $1.6$ and $2.2$.

For ALPs, we considered both continuous reactor-photon production and
monochromatic production from nuclear magnetic transitions. The
single-coupling searches involving $g_{a\gamma\gamma}$ and $g_{aee}$
display nearly mass-independent sensitivities in their long-lived,
low-mass regimes, while finite lifetimes generate additional structure at
larger masses. In particular, the opening of
$a\rightarrow e^+e^-$ at $m_a=2m_e$ produces a characteristic decay
region in the electron-coupling sensitivity. Searches based on
$p(n,\gamma)d$ production provide complementary access to the products
$|g_{a\gamma\gamma}g_{ann}^{(1)}|$ and
$|g_{aee}g_{ann}^{(1)}|$, with a sharp loss of sensitivity as
$m_a$ approaches the $2.230~{\rm MeV}$ nuclear-transition endpoint.

Compared with current experimental limits only (laboratory, reactor and
terrestrial solar-ALP searches), the single-coupling projections lie inside
regions that are already excluded; the closest approach is for
$g_{a\gamma\gamma}$ at $m_a\simeq0.5~{\rm MeV}$, where CONUS100 is a factor
$\simeq1.8$ above the NEON limit.  The smallest coupling products, which
correspond to long-lived ALPs, are excluded by Borexino.  Two regions,
however, are not covered by published experimental limits: the
$a\to e^+e^-$ window of $|g_{aee}g_{ann}^{(1)}|$ between $2m_e$ and
$2.230~{\rm MeV}$, reaching $1.3\times10^{-15}$ for
CONUS100, and, at MeV masses, the region of the
$(g_{a\gamma\gamma},g_{ann}^{(1)})$ and $(g_{aee},g_{ann}^{(1)})$ planes in
which the ALP decays between the Sun and the Earth but survives the
$20~{\rm m}$ reactor baseline.  Astrophysical constraints, which were not
considered here, are expected to be relevant in the same regions.
Both regions require a readout window extending to MeV energies, as
assumed here for CONUS100: with its present $0.6~\mathrm{MeV}$ DAQ range,
CONUS+ is sensitive only to the single-coupling channels below
$0.6~\mathrm{MeV}$, which are already excluded.

Across all channels, the CONUS100 benchmark benefits substantially from
the increase in target mass and integrated exposure. Its largest gain is
not simply a rescaling of the CONUS+ result: the improvement depends on
the signal-production mechanism, the detector channel, the ALP lifetime,
and, for mixed-coupling searches, the separate coupling dependence of
propagation and detection. The resulting projections therefore illustrate
the broad physics reach that a larger reactor-based germanium experiment
could achieve for weakly coupled particles.

The present sensitivities should nevertheless be interpreted as idealized
projections. They use a common flat background benchmark of
$10~{\rm DRU}$, simplified counting statistics, and unit signal efficiency
within the corresponding analysis regions. A future experimental analysis
should incorporate the measured energy-dependent background, detector
response and containment efficiencies, reactor-on and reactor-off data,
and a likelihood treatment of systematic and nuisance parameters.
Implementing these ingredients will be necessary to convert the present
physics reach estimates into detector-level exclusion limits.

Overall, the results show that reactor experiments optimized for low-energy neutrino measurements can also probe light dark-sector particles.  With its present DAQ, CONUS+ can already improve on the TEXONO dark photon limit by a factor $\simeq1.6$, while the larger exposure and extended readout envisaged for CONUS100 give access to unexplored ALP parameter space at MeV masses.

\newpage

\acknowledgments
We are grateful to the CONUS Collaboration for useful discussions.
FSQ is supported by the Humboldt Foundation (Friedrich Wilhelm Bessel Research Award), ANID - Millennium Science Initiative Program - ICN2019\_044, IIF-FINEP grant 213/2024, FAPESP Grants 2021/14335-0 and 2021/00449-4, and CNPq grants 307130/2021-5, 403521/2024-6, 309465/2026-5, and 446711/2025-0. LA acknowledges support from the Coordenação de Aperfeiçoamento de Pessoal de Nível Superior (CAPES) under grants 88887.827404/2023-00 and 88881.128190/2025-01. LA also thanks the Department of Physics at the University of Oslo for the hospitality, support and useful discussions during the completion of this work. We also thank IIP for the local cluster {\it bulletcluster}, which was instrumental in this work. We acknowledge the use of Claude (Anthropic) for assistance in checking the numerical codes and in the language editing of the manuscript; the authors take full responsibility for the content of this work.



\bibliographystyle{JHEP}
\bibliography{biblio.bib}

\end{document}